\PassOptionsToPackage{unicode}{hyperref}
\PassOptionsToPackage{hyphens}{url}
\PassOptionsToPackage{dvipsnames,svgnames,x11names}{xcolor}
\documentclass[12pt]{article}

\usepackage{amsthm}
\usepackage{bbm}
\usepackage{bm}
\usepackage{subcaption}
\usepackage{soul}
\usepackage{algorithm}
\usepackage{algpseudocode}
\usepackage{graphicx}
\usepackage{multirow}

\newtheorem{prop}{Proposition}

\newtheorem{theorem}{Theorem}[section]

\newtheorem{assumption}{Assumption}

\renewcommand{\baselinestretch}{1}
\newtheorem{corollary_1}[theorem]{Corollary}
\usepackage{wrapfig}
\usepackage{lipsum}
\usepackage{comment}
\usepackage{amsmath,amssymb}
\usepackage{iftex}
\ifPDFTeX
  \usepackage[T1]{fontenc}
  \usepackage[utf8]{inputenc}
  \usepackage{textcomp} 
\else 
  \usepackage{unicode-math}
  \defaultfontfeatures{Scale=MatchLowercase}
  \defaultfontfeatures[\rmfamily]{Ligatures=TeX,Scale=1}
\fi
\usepackage{lmodern}
\ifPDFTeX\else  
\fi
\IfFileExists{upquote.sty}{\usepackage{upquote}}{}
\IfFileExists{microtype.sty}{
  \usepackage[]{microtype}
  \UseMicrotypeSet[protrusion]{basicmath} 
}{}
\makeatletter
\@ifundefined{KOMAClassName}{
  \IfFileExists{parskip.sty}{%
    \usepackage{parskip}
  }{
    \setlength{\parindent}{0pt}
    \setlength{\parskip}{6pt plus 2pt minus 1pt}}
}{
  \KOMAoptions{parskip=half}}
\makeatother
\usepackage{xcolor}
\makeatletter
\ifx\paragraph\undefined\else
  \let\oldparagraph\paragraph
  \renewcommand{\paragraph}{
    \@ifstar
      \xxxParagraphStar
      \xxxParagraphNoStar
  }
  \newcommand{\xxxParagraphStar}[1]{\oldparagraph*{#1}\mbox{}}
  \newcommand{\xxxParagraphNoStar}[1]{\oldparagraph{#1}\mbox{}}
\fi
\ifx\subparagraph\undefined\else
  \let\oldsubparagraph\subparagraph
  \renewcommand{\subparagraph}{
    \@ifstar
      \xxxSubParagraphStar
      \xxxSubParagraphNoStar
  }
  \newcommand{\xxxSubParagraphStar}[1]{\oldsubparagraph*{#1}\mbox{}}
  \newcommand{\xxxSubParagraphNoStar}[1]{\oldsubparagraph{#1}\mbox{}}
\fi
\makeatother

\usepackage{longtable,booktabs,array}
\usepackage{calc} 
\usepackage{etoolbox}
\makeatletter
\patchcmd\longtable{\par}{\if@noskipsec\mbox{}\fi\par}{}{}
\makeatother
\IfFileExists{footnotehyper.sty}{\usepackage{footnotehyper}}{\usepackage{footnote}}
\makesavenoteenv{longtable}
\usepackage{graphicx}
\makeatletter
\def\maxwidth{\ifdim\Gin@nat@width>\linewidth\linewidth\else\Gin@nat@width\fi}
\def\maxheight{\ifdim\Gin@nat@height>\textheight\textheight\else\Gin@nat@height\fi}
\makeatother
\setkeys{Gin}{width=\maxwidth,height=\maxheight,keepaspectratio}
\makeatletter
\def\fps@figure{htbp}
\makeatother

\makeatletter
\@ifpackageloaded{caption}{}{\usepackage{caption}}
\AtBeginDocument{%
\ifdefined\contentsname
  \renewcommand*\contentsname{Table of contents}
\else
  \newcommand\contentsname{Table of contents}
\fi
\ifdefined\listfigurename
  \renewcommand*\listfigurename{List of Figures}
\else
  \newcommand\listfigurename{List of Figures}
\fi
\ifdefined\listtablename
  \renewcommand*\listtablename{List of Tables}
\else
  \newcommand\listtablename{List of Tables}
\fi
\ifdefined\figurename
  \renewcommand*\figurename{Figure}
\else
  \newcommand\figurename{Figure}
\fi
\ifdefined\tablename
  \renewcommand*\tablename{Table}
\else
  \newcommand\tablename{Table}
\fi
}
\@ifpackageloaded{float}{}{\usepackage{float}}
\floatstyle{ruled}
\@ifundefined{c@chapter}{\newfloat{codelisting}{h}{lop}}{\newfloat{codelisting}{h}{lop}[chapter]}
\floatname{codelisting}{Listing}

\makeatother
\makeatletter
\@ifpackageloaded{caption}{}{\usepackage{caption}}
\@ifpackageloaded{subcaption}{}{\usepackage{subcaption}}
\makeatother

\ifLuaTeX
  \usepackage{selnolig}  
\fi
\usepackage[]{natbib}
\usepackage{bookmark}

\IfFileExists{xurl.sty}{\usepackage{xurl}}{} 
\hypersetup{
  pdftitle={Title},
  pdfauthor={Author 1; Author 2},
  pdfkeywords={3 to 6 keywords, that do not appear in the title},
  colorlinks=true,
  linkcolor={blue},
  filecolor={Maroon},
  citecolor={Blue},
  urlcolor={Blue},
  pdfcreator={LaTeX via pandoc}}

\newcommand{\anon}{1}

\DeclareMathOperator*{\argmax}{arg\,max}

\begin{document}

\def\spacingset#1{\renewcommand{\baselinestretch}%
{#1}\small\normalsize} \spacingset{1}


\if1\anon
{
  \title{\bf Inferring Diffusion Structures of Heterogeneous Network Cascade}
  \author{Yubai Yuan  \thanks{Co-first authors}
  \thanks{
    Corresponding author, email:\textit{yvy5509@psu.edu}}\\
    Department of Statistics, The Pennsylvania State University\\
    and \\
    Siyu Huang \footnotemark[1] \\
    Department of Statistics, The Pennsylvania State University\\
    and \\
    Abdul Basit Adeel \\
    Department of Sociology and Criminology, The Pennsylvania State University
    }
 \date{}
  \maketitle
 \fi
 }

\if0\anon
{
  \bigskip
  \bigskip
  \bigskip
  \begin{center}
    {\LARGE\bf Title}
\end{center}
  \medskip
} \fi

\bigskip
\begin{abstract}

Network cascade refers to diffusion processes in which outcomes changes within part of an interconnected population trigger a sequence of changes across the entire network. These cascades are governed by underlying diffusion networks, which are often latent. Inferring such networks is critical for understanding cascade pathways, uncovering Granger causality of interaction mechanisms among individuals, and enabling tasks such as forecasting or maximizing information propagation. In this project, we propose a novel double mixture directed graph model for inferring multi-layer diffusion networks from cascade data. The proposed model represents cascade pathways as a mixture of diffusion networks across different layers, effectively capturing the strong heterogeneity present in real-world cascades. Additionally, the model imposes layer-specific structural constraints, enabling diffusion networks at different layers to capture complementary cascading patterns at the population level. A key advantage of our model is its convex formulation, which allows us to establish both statistical and computational guarantees for the resulting diffusion network estimates. We conduct extensive simulation studies to demonstrate the model’s performance in recovering diverse diffusion structures. Finally, we apply the proposed method to analyze cascades of research topic in the social sciences across U.S. universities, revealing the underlying diffusion networks of research topic propagation among institutions.
\end{abstract}

\noindent%

\vfill

\newpage
\spacingset{1.8} 

\section{Introduction}\label{sec-intro}

Cascade refers to the diffusion processes that outcomes changes happening in a part of an interconnected population leads to a series of sequential changes throughout the entire population. In recent years, there is a surging interests and efforts to understand and model the cascade mechanisms due to that it motivates many significant research topics from different disciplines, and generalizes a broad range of network-related applications in real world. Cascading behavior arise from social influence \citep{french1959bases, friedkin2011social}, propagation of information on social media \citep{romero2011differences}, adoption of new products and technical innovations \citep{aral2012identifying}, diffusion of policy and social norms \citep{shipan2008mechanisms}, viral marketing \citep{leskovec2007dynamics}, and contagion of infectious diseases \citep{keeling2005networks}. Cascade dynamics is also common in various micro-level network systems such as protein and gene-interaction networks \citep{maslov2002specificity} and brain network \citep{plenz2007organizing}.  


Understanding the cascade processes boils down to recover the underlying diffusion networks that govern the propagation structures of cascades. However, the diffusion networks are often hidden. For example, in the case of infectious diseases, the exact cascade pathway is typically missing and 
it is unclear who transmits the diseases to an infected individual. On the other hand, we can often observe the results of the diffusion process, i.e., the timestamps when each individual is activated in the cascade. The activation time lag between two individuals implies their distance on the diffusion network. Therefore, by collecting and properly modeling the activation 
time lags from repeated cascades over the population, one can infer the transmission patterns among the individuals and recover the diffusion network structure. Following this intuition, a variety of directed graphical models are developed to model cascade data and infer latent diffusion networks \citep{gomez2012inferring, gomez2013structure, rodriguez2014uncovering, daneshmand2014estimating, kempe2003maximizing, gomez2013modeling}. In general, the existing methods model the individual activation time as continuous random variable and construct the likelihood of observed cascades as Markov processes over the latent diffusion networks. The diffusion network estimated as a directed and weighted graph, where the direction and weight on the edge represents the sender/receiver role and the transmission rate between the corresponding individuals. 

\begin{figure}[h]
    \begin{minipage}[b]{0.32\linewidth}
        \centering
        \includegraphics[width=\textwidth]{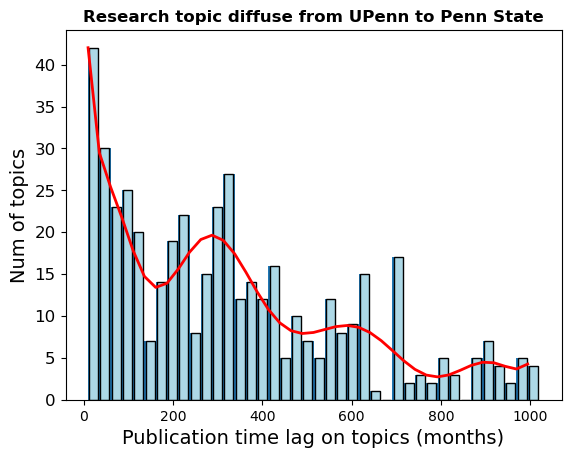}
    \end{minipage}
    \begin{minipage}[b]{0.32\linewidth}
        \centering
        \includegraphics[width=\textwidth]{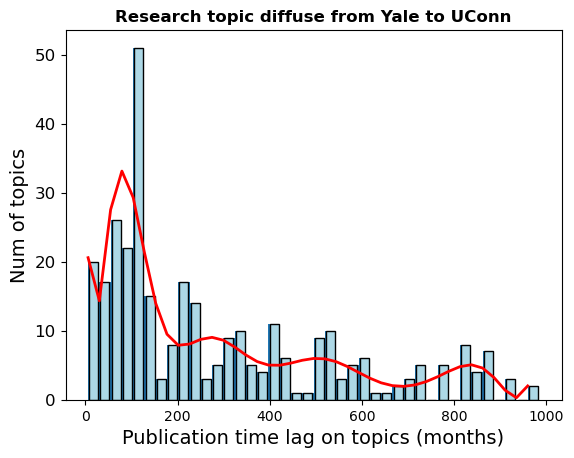}
    \end{minipage}
     \begin{minipage}[b]{0.32\linewidth}
        \centering
        \includegraphics[width=\textwidth]{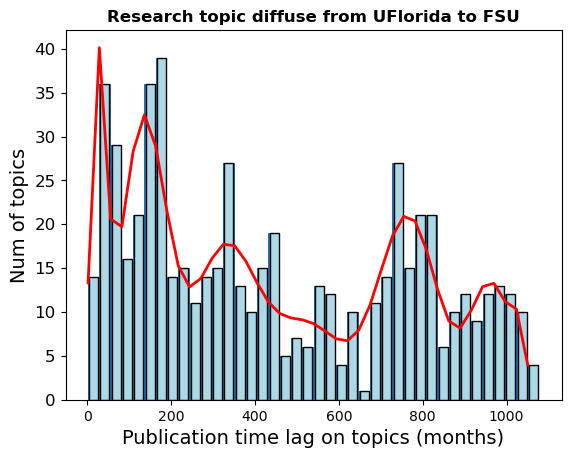}
    \end{minipage}
\caption{{\small Publication time lags of three pairs of neighboring universities} across research topics.}
    \label{cascade}
\end{figure}

Most of existing methods consider the scenario of homogeneous cascade where all observed cascades follow the same diffusion pattern. While this assumption may hold in specific applications, the real-world cascades typically exhibit strong heterogeneous diffusion patterns, especially in information cascade and epidemiology \citep{wang2014mmrate, yu2020estimation, du2013scalable}. One motivating example is study the research topic diffusion among U.S. universities. We generate about 3,000 important topics in sociology from 29,725 articles published in 23 top sociology journals. For each topic, we define a university's "activation time" as the publication date of the first article on the topic whose first author is affiliated with that university. Figure \ref{cascade} illustrates the distribution of activation time lags between three pairs of geographically close universities: UPenn and Penn State, Yale and UConn, and UFlorida and FSU. Despite the strong geographical relationship for each pair of universities, the multimodal patterns in the time lag distributions suggest significant heterogeneity in topic diffusion. 

One major source of cascade heterogeneity is due to multi-type relations such that cascades vary on propagation speeds and scales when diffusing via different types of relation \citep{klovdahl1994social, oh2008examining}. To account for the cascade-dependent diffusion patterns, several cascade models are proposed \citep{wang2014mmrate, yu2020estimation} to first cluster cascades into several groups and then estimate a diffusion network for each group of diffusion pattern. However, the existing methods are still limited in capturing diversity of diffusion patterns due to the constraints on the number of cascade groups. In addition, these methods require additional cascade labels for the clustering. More importantly, the existing methods fail to capture the individual-dependent cascade heterogeneity \citep{watts2002simple, duan2009informational}. For example, the spreading pattern and speed of the messages in social media heavily depend on users' activity status. When being active, an user will respond more instantly to interesting message and accelerate the information cascade compared with the inactive status \citep{du2012learning}. For the volatility cascade in financial markets, the volatility diffusion patterns depend on time horizons of active agents on the market, which varies across financial events \citep{zumbach2001heterogeneous}. In real-world cascades, the individual status affects both how the individual is activated and how the individual transmits the cascade to the downstream population. Collectively, the global diffusion pattern can exhibit heterogeneity at combinatorial complexity as the population size grows. Additionally, unlike observable and static cascade information, the individual status is typically infeasible to collect and constantly changes across cascade events. The above individual-level heterogeneity posts significant challenges for identifying and discriminating the basic diffusion patterns from cascade data.

In this paper, we propose a novel double mixture graphical model to infer latent diffusion networks from cascade data. The proposed model represents each observed cascade as a mixture of Markov processes over multi-layer diffusion networks. This formulation enables principled modeling of cascade heterogeneity and allows for the recovery of latent diffusion networks even when the number of possible diffusion patterns grows exponentially with the population size. To identify diffusion networks at different layers, we introduce layer-specific structural regularization, inspired by the rank-sparsity incoherence condition in matrix decomposition \citep{chandrasekaran2011rank}. This structural regularization also allows each layer to capture cascade transmission driven by different types of relationships among individuals. Additionally, our method does not require external information on individual statuses during the cascade, which are treated as missing data and inferred in a data-adaptive manner. We develop a regularized high-dimensional EM-type algorithm to estimate the diffusion matrices, allowing the network dimensions to diverge. Each parameter update in the EM algorithm can be efficiently solved as convex optimization. Leveraging the convexity of the model formulation, we establish non-asymptotic results for both the computational and statistical convergence of the network estimations. Unlike existing EM algorithm analyses in vector parameter space \citep{10.1214/16-AOS1435, wang2014high}, our theoretical results are derived for matrix-valued parameters and can incorporate low-dimensional structures such as sparsity and low rank. Furthermore, our convergence guarantees do not require specific initialization conditions.


\vspace{-5mm}

\section{Directed graph model for cascade data}\label{sec-2}

Consider a population with $N$ units $\mathcal{V} = \{1,\cdots,N\}$, and each unit has two cascade outcomes: activated (infected) and inactivated (non-infected). The cascade outcome does not change once a unit is activated. A cascade is a $N$-dimensional temporal record $\bm{t} = (t_1,t_2,\cdots,t_N)$, where $t_i$ is the infection time of the $i$-th unit. We observe cascade within a finite window of length $T$, i.e., $t_i \in [t_0, t_0 + T]$ where $t_0 := \min_{1\leq i \leq N}\{t_i\}$ is the activation time of the source node. Denote the activation time of units not infected within the observation window as $t_0 + T$. Without loss of generality, we assume $t_0 = 0$ in this paper.


A cascade initializes at a source node at $t_0$ and propagates via a \textit{directed} diffusion network $\mathcal{E}\in \mathrm{R}_{+}^{N\times N}=(e_{ij})$ over $\mathcal{V}$ where $\mathcal{E}$ is unobserved. As the building block of the cascade process, the directed graph models \citep{gomez2012inferring, rodriguez2014uncovering, gomez2013modeling} formulate the cascade transmission as survival data.  
Specifically, given node $j$ being infected at $t_j$, the transmission $f(t_i\mid t_j,{e}_{ji})$ describing the likelihood of node $i$ being infected by $j$ at $t_i$ where $t_i > t_j$. The directed and weighted edge $e_{ji}$ represents how fast cascade propagates from node $j$ to $i$. The cumulative infection probability is $F(t_i\mid t_j,e_{ji}) = \int_{t_j}^{t_i}f(s\mid t_j,e_{ji}) \mathrm{d}s$, with survival function and hazard function being
\begin{align}\label{eq_1}
S(t_i\mid t_j,e_{ji}) = 1 - F(t_i\mid t_j,e_{ji});\; H(t_i\mid t_j,e_{ji}) = \frac{f(t_i\mid t_j,e_{ji})}{S(t_i\mid t_j,e_{ji})},
\end{align}
Intuitively, $H(t_i\mid t_j,e_{ji})$ measures the probability of node $i$ get infected by node $j$ within time window $[t_i, t_i + \Delta t ]$ given that node $i$ is not infected by node $j$ before $t_i$. 
The shape of the transmission probability depends on the parametric form of hazard function, which typically use three well-known models: exponential (Exp), power-law (Pow), and Rayleigh (Ray) models. Specifically, the monotone Exp model $H(t_i\mid t_j,e_{ji}) = e_{ji}\mathbbm{1}(t_i > t_j)$ and Pow model $H(t_i\mid t_j,e_{ji}) = \frac{e_{ji}}{t_i -t_j}\mathbbm{1}(t_i - t_j > \delta)$ are previously used in modeling diffusion on social networks \citep{myers2010convexity}, and the Ray model $H(t_i\mid t_j,e_{ji}) = e_{ji}(t_i - t_j)\mathbbm{1}(t_i > t_j)$ is used in epidemiology due to its adaption to model rise-then-fall pattern in contagion.  
A larger $e_{ji}$ implies a faster transmission of cascade from node $j$ to node $i$.
Notice that $\mathcal{E}$ can be asymmetric, i.e., $e_{ij}\neq e_{ji}$ when node $i$ and $j$ are different in the capacity of launching transmission. 
The conditional probability of activating node $i$ at $t_i$ given activated nodes $\{j: t_j<t_i\}$ is formulated as the independent cascade model \citep{kempe2003maximizing}, i.e., node $i$ gets activated once the first parent activates it. The conditional probability of activation is 
\begin{align}\label{eq_5_5}
\mathbb{P}(t_i \mid \{t_j:t_j<t_i\}) = \mathbb{P}_{I}(t_i; \mathcal{E}_{\cdot i}): = \sum_{j:t_j<t_i}f(t_i\mid t_j, e_{ji})\prod_{k: t_k<t_i, k\neq j}S(t_i\mid t_k, e_{ki}),
\end{align}
where each term within the above summation corresponds to the probability of an activated node to be the first parent. On the other hand, if the node $i$ is not activated by all activated nodes within observation window $T$, the corresponding probability is 
\begin{align}\label{eq_5_6}
\mathbb{P}(t_i = T \mid \{t_j:t_j<t_i\}) = \mathbb{P}_{U}(T; \mathcal{E}_{\cdot i}): = \prod_{j:t_j<T}S(T \mid t_j, e_{ji}).
\end{align}
Notice that the activation time $t_i$ only depends on the $i$-th column of the adjacent matrix of the diffusion network $\mathcal{E}$. The directed graph model assumes that node-wise activations are independent conditioning on the parents. Therefore, the likelihood of an observed cascade can be factorized over nodes as
\begin{align}\label{eq_6}
\mathbb{P}(\bm{t};\mathcal{E}) \! = \!  \prod_{i=1}^N\mathbb{P}(t_i\mid \{t_j: t_j < t_i\};\mathcal{E}_{\cdot i} ) \!=\!  \prod_{i=1}^N \big\{ \mathbbm{1}( t_i<T) \mathbb{P}_{I}(t_i; \mathcal{E}_{\cdot i}) \!+\! \mathbbm{1}( t_i\geq T) \mathbb{P}_{U}(T; \mathcal{E}_{\cdot i})\big\}, 
\end{align}
and the full likelihood on independent cascade samples $\{ \bm{t}^{(c)} \}_{c=1}^C $ over $\mathcal{E}$ is
\begin{align}\label{eq_7}
\prod_{c=1}^C\mathbb{P}(\bm{t}^{(c)} ; \mathcal{E} )\!=\!\prod_{c=1}^C  \!\prod_{i:t^c_i\leq T} \Big[ \sum_{k:t^c_k < t^c_i} H(t^c_i\mid t^c_k, e_{ki}) \!\times\! \!\!\prod_{j: t^c_j < t^c_i}S(t^c_i\mid t^c_j, e_{ji}) \!\times\!\!\!\! \prod_{m: t^c_m \geq T }S(T \mid t^c_i, e_{im})\Big].
\end{align}
Notice that each cascade sample induces a directed acyclic graph (DAG) with support on the diffusion network $\mathcal{E}$. The local Markov property of DAG allows one to decompose the likelihood and reduce the computational complexity for inferring diffusion network.

\vspace{-5mm}
\section{Method}\label{sec-3}

In real applications, the diffusion patterns of cascades exhibit strong heterogeneity. In addition, there typically exists multiple layers in the diffusion networks, where layer reflects different relations and has different diffusion patterns. Furthermore, a cascade can propagate alternatively across different layers via inter-layer interactions. To model the cross-layer cascading behavior, we propose a mixture cascade generative process in this section.  

\vspace{-5mm}
\subsection{Double mixture directed graph model}

Consider two diffusion networks $\bm{\Theta}\in \mathrm{R}_{+}^{N\times N}$ and $\bm{\Psi}\in \mathrm{R}_{+}^{N\times N}$ among the same population $\mathcal{V}$. 
We introduce diffusion network indicators $Z_{i}^c \in \{0,1\}$ such that
\begin{align*}
    Z_{i}^c \overset{\text{i.i.d}}{\sim} \text{Bern}(\pi_i),\; i = 1,\cdots, N, \; c = 1,\cdots, C.
\end{align*}
The random variable $Z_{i}^c$ indicates through which diffusion network node $i$ engages with the $c$-th cascade. For the $c$-th cascade, $Z_{i}^c = 1$ if node $i$ is activated via 
diffusion network $\bm{\Theta}$, and $Z_{i}^c = 0$ if via network $\bm{\Psi}$.
The parameter $\bm{\pi}  = \{\pi_i\}_{i=1}^N \in [0,1]^N$ are the node-wise probabilities of activation network. Based on network indicators $\{Z_i^c\}_{i=1}^N$, the diffusion network  of the $c$-th cascade $\mathcal{E}^c$ is: 
\begin{align}\label{dou_mix}
    \mathcal{E}^c = \big( Z_1^c\bm{\Theta}_{\cdot 1} + (1-Z_1^c)\bm{\Psi}_{\cdot 1},\cdots, Z_N^c\bm{\Theta}_{\cdot N} + (1-Z_N^c)\bm{\Psi}_{\cdot N}\big),\; c = 1,\cdots, C.
\end{align}
Therefore, $\mathcal{E}^c$ is column-wise mixture of $\bm{\Theta}$ and $\bm{\Psi}$. Different to traditional mixture model with fixed node-wise membership, the proposed mixture model allows the node-wise network membership to vary across cascade, i.e., $\{Z_i^c\}$ can be different in terms of both $i$ and $c$. Compared with an additive model $\mathcal{E}^c = \bm{\Theta} + \bm{\Psi}$ where diffusion network is still homogeneous for all cascades, the column-wise mixture in (\ref{dou_mix}) can generate up to $2^N$ different diffusion patterns, which grows exponentially as network size increases. 
Therefore, the proposed model provides a principled flexibility to capture the strong heterogeneity in observed cascades, and captures the cross-layer propagation process. 

The likelihood of the proposed model can be explicitly formulated. Following (\ref{eq_5_5}), the probability of node $i$ being activated at $t_i<T$ conditioning on $\mathcal{E}^c$ in the $c$-th cascade is: 
\begin{align}\label{eq_6_5}
\mathbb{P}(t^c_i\mid \{t^c_j: t^c_j < t^c_i\}, Z^c_i; \mathcal{E}^c)  = \big[\mathbb{P}_{I}(t^c_i; \bm{\Theta}_{\cdot i})\big]^{Z^c_i} \times \big[\mathbb{P}_{I}(t^c_i; \bm{\Psi}_{\cdot i})\big]^{1 - Z^c_i},
\end{align}
and following (\ref{eq_5_6}) the probability for node $i$ not being activated is
\begin{align}\label{eq_6_6}
    \mathbb{P}(T \mid \{t^c_j: t^c_j < T\}, Z^c_i; \mathcal{E}^c)  = \big[\mathbb{P}_{U}(T; \bm{\Theta}_{\cdot i})\big]^{ Z^c_i} \times \big[\mathbb{P}_{U}(T; \bm{\Psi}_{\cdot i})\big]^{1 - Z^c_i},
\end{align}
where $\mathbb{P}_I(\cdot)$ and $\mathbb{P}_U(\cdot)$ are defined in (\ref{eq_5_5}) and (\ref{eq_5_6}).  
Denote $\bm{Z} = \{Z_i^c\}_{i=1}^N,\;c= 1,\cdots, C$, the joint distribution of network indicators is 
\begin{align}\label{eq_9}
\mathbb{P}(\bm{Z}) = \prod_{c=1}^C\prod_{i=1}^N\mathbb{P}(Z^c_{i}) = \prod_{c=1}^C\prod_{i = 1}^N\pi_i^{Z^c_i}(1 - \pi_i)^{1-Z^c_i}.
\end{align} 
Let $\bm{\Omega} = ( \bm{\pi}, \bm{\Theta}, \bm{\Psi})$ be model parameters, then the joint likelihood of cascade samples and network indicators is 
\begin{align}\label{eq_9_2}
 \prod_{c=1}^C \mathbb{P}(\bm{t}^{(c)},\bm{Z}; \bm{\Omega}) = \prod_{c=1}^C \Big\{  &\prod_{i:t^c_i\leq T}   \big[ \pi_i\mathbb{P}_{I}(t^c_i; \bm{\Theta}_{\cdot i})\big]^{Z^c_i}  \big[(1-\pi_i) \mathbb{P}_{I}(t^c_i; \bm{\Psi}_{\cdot i})\big]^{1 - Z^c_i}  \nonumber \\ \times 
  &\prod_{j: t^c_j \geq T } \big[ \pi_j\mathbb{P}_{U}(T; \bm{\Theta}_{\cdot i})\big]^{ Z^c_i} \big[(1-\pi_j)\mathbb{P}_{U}(T; \bm{\Psi}_{\cdot i})\big]^{1 - Z^c_i} \Big\},
\end{align}
and the marginal likelihood of cascade samples is
\begin{align}\label{eq_10}
    \prod_{c=1}^C \mathbb{P}(\bm{t}^{(c)};\bm{\Omega}) = \prod_{c=1}^C \Big\{ \prod_{i:t^c_i\leq T}  \Big[ 
    \pi_i \mathbb{P}_{I}(t^c_i; \bm{\Theta}_{\cdot i}) + (1-\pi_i) \mathbb{P}_{I}(t^c_i; \bm{\Psi}_{\cdot i})
    \Big]\times \\  \nonumber  \prod_{j: t^c_j \geq T }\Big[  \pi_i\mathbb{P}_{U}(T; \bm{\Theta}_{\cdot i}) + (1-\pi_i) \mathbb{P}_{U}(T; \bm{\Psi}_{\cdot i})   \Big]  \Big\}
\end{align}
Based on (\ref{eq_9_2}) and (\ref{eq_10}), the posterior probability of network indicators $\{Z_i^c\}$ can be derived in an explicit form as: 
\begin{align}\label{E_step}
    \hat{\pi}^c_i := \mathbb{P}(Z_i^c = 1 \mid \bm{t}^{(c)}) = 
    \begin{cases}
     \frac{ \pi_i \mathbb{P}_{I}(t^c_i; \bm{\Theta}_{\cdot i}) }{ \pi_i \mathbb{P}_{I}(t^c_i; \bm{\Theta}_{\cdot i}) 
     + (1-\pi_i) \mathbb{P}_{I}(t^c_i; \bm{\Psi}_{\cdot i})},\; \text{if}\; t_i^c <T \\
     \frac{\pi_i  \mathbb{P}_{U}(t^c_i; \bm{\Theta}_{\cdot i}) }{  \pi_i \mathbb{P}_{U}(t^c_i; \bm{\Theta}_{\cdot i})  +   (1-\pi_i) \mathbb{P}_{U}(t^c_i; \bm{\Psi}_{\cdot i}) },\; \text{if}\; t_i^c \geq T
    \end{cases}. 
\end{align}

\subsection{Model identification}

We establish the identifiability of the proposed mixture model. 
Denote $\mathcal{R}^{\Theta}_i := \{ j \in \{1,\cdots, N\} \mid \bm{\Theta}_{ji} > 0\}$ and $\mathcal{R}^{\Psi}_i := \{ j \in \{1,\cdots, N\} \mid \bm{\Psi}_{ji} > 0\}$ as the sets of parents of node $i$ in $\bm{\Theta}$ and $\bm{\Psi}$, respectively. Then $\bm{t}_{\mathcal{R}_i}\in [0,T]^{|\mathcal{R}^{\Theta}_i \cup \mathcal{R}^{\Psi}_i|}: = \{ t_j \mid j \in \mathcal{R}^{\Theta}_i \cup \mathcal{R}^{\Psi}_i \}$ as the activation times of parent nodes.  
We have the identification for model in (\ref{eq_10}) as follows.
\begin{prop}
For each node $i,\; i =1,\cdots, N$, assume that 1) $\| \bm{\Theta}_{\cdot i} \|_1 + \| \bm{\Psi}_{\cdot i} \|_1 >0$ and there exist $j\neq i$ such that $\bm{\Theta}_{ji} \neq  \bm{\Psi}_{ji}$, and 2) survival function satisfies $S(t_i \mid t_j, \theta_{ji}) = \exp\{ \theta_{ji}h(t_i - t_j)\}$ for some differentiable function $h(\cdot)$. 
The parameters $(\pi_i,\bm{\Theta}_{\cdot i},\bm{\Psi}_{\cdot i} )$ associated with node $i$ in (\ref{eq_10}) are identifiable, i.e., if 
\begin{align*}
    &\pi_i \mathbb{P}_{I}(t; \bm{\Theta}_{\cdot i}) + (1-\pi_i) \mathbb{P}_{I}(t; \bm{\Psi}_{\cdot i}) = \tilde{\pi}_i {\mathbb{P}}_{I}(t; \tilde{\bm{\Theta}}_{\cdot i}) + (1-\tilde{\pi}_i) \mathbb{P}_{I}(t; \tilde{\bm{\Psi}}_{\cdot i}), \;\text{or} \\
    &\pi_i \mathbb{P}_{U}(t; \bm{\Theta}_{\cdot i}) + (1-\pi_i) \mathbb{P}_{U}(t; \bm{\Psi}_{\cdot i}) = \tilde{\pi}_i {\mathbb{P}}_{U}(t; \tilde{\bm{\Theta}}_{\cdot i}) + (1-\tilde{\pi}_i) \mathbb{P}_{U}(t; \tilde{\bm{\Psi}}_{\cdot i})
\end{align*}
for any $t_i$ and $\bm{t}_{\mathcal{R}_i}$, then $\pi_i = \tilde{\pi}_i$, ${\bm{\Theta}}_{\cdot i} = \tilde{\bm{\Theta}}_{\cdot i}$, and ${\bm{\Psi}}_{\cdot i} = \tilde{\bm{\Psi}}_{\cdot i}$. 
\end{prop}
\textit{Remark}: 
Assumption 2 requires that the parameter and time shaping function are separable in hazard function, which can be satisfied by the popular survival models including Exp model, Pow model, Ray model, and other additive models of information propagation such as kernelized \citep{du2012learning} and featured-enhanced hazard functions \citep{wang2012feature}.   

Proposition 1 guarantees that parameters $\{\pi_i\}_{i=1}^N$ are identifiable, and the diffusion networks $\bm{\Theta}$ and $\bm{\Psi}$ are column-wise identifiable. However, similar to the labeling non-identifiability issue in finite mixture model \citep{kim2015empirical}, $\bm{\Theta}$ and $\bm{\Psi}$ may still not be globally identifiable without structural constraints due to column permutation.  Specifically, the data distribution does not change if we swap ${\bm{\Theta}}_{\cdot i}$ and ${\bm{\Psi}}_{\cdot i}$ in ${\bm{\Theta}}$ and ${\bm{\Psi}}$ with other columns fixed. 

Motivated from real applications, one can interpret $\bm{\Theta}$ as
diffusion over an observed structural network $\bm{A}\in \{0,1\}^{N\times N}$. Therefore, we can impose support constraints on $\bm{\Theta}$ as 
\begin{align*}
    \bm{\Theta}_{ij} \geq 0 \;\text{if}\; \bm{A}_{ij} =1;\; \bm{\Theta}_{ij} = 0 \;\text{if}\; \bm{A}_{ij} =0.
\end{align*}
Due to the sparsity nature of many real-world networks, the support constraint also implicitly imposes the sparsity on $\bm{\Theta}$. 
On the other hand, one can interpret $\bm{\Psi}$ as the diffusion over latent relations among the units that are not captured by the structural network $\bm{A}$. The magnitude of $\bm{\Psi}_{ij}$ reflects the distance between unit $i$ and $j$ in terms of their latent factors. It has been found that large-scale networks typically has or can be approximated by low-rank structure \citep{udell2019big, liben2007link, menon2011link} due to that high-dimensional latent factors can be always embeded into a low dimensional latent space that preserve original distances \citep{udell2019big}. Therefore, we impose low-rank structure on $\bm{\Psi}$ as
\begin{align*}
    \text{rank}(\bm{\Psi})\leq r, \; 1\leq r << N.
\end{align*}

Imposing above support and low-rank structure allows $\bm{\Theta}$ and $\bm{\Psi}$ to capture complementary diffusion patterns driven by different types of relations. In addition, the structure constraints can solve the above column-wise permutation issue.
We follow \citep{chandrasekaran2011rank} to introduce the subspace of matrix $\boldsymbol{\Lambda_1}(\bm{\Theta} )=\left\{N \in \mathbb{R}^{N \times N} \mid \operatorname{support}(N) \subseteq \bm{A}\right\}$. We perform SVD on $\bm{\Psi} = \bm{U}\bm{\Sigma}\bm{V}^{\top}$ where $\bm{U},\bm{V} \in \mathbb{R}^{N\times r}$ and $r$ is the rank of $\bm{\Psi}$. Then define another subspace of matrix as $\boldsymbol{\Lambda_2}(\bm{\Psi}) =\left\{\bm{U} X^{\top}+Y \bm{V}^{\top} \mid X, Y \in \mathbb{R}^{N \times k}\right\}$. We have:
\begin{prop}
 Under assumptions in Proposition 1, $\bm{\Theta}$ and $\bm{\Psi}$ are identifiable if:
 \begin{align}\label{eq_11}
     \max _{N \in \boldsymbol{\Lambda_1}(\bm{\Theta}),\|N\|_{\infty} \leq 1}\|N\|_2 \times \max _{N \in \boldsymbol{\Lambda_2}(\bm{\Psi}),\|N\|_2 \leq 1}\|N\|_{\infty} < 1,
 \end{align}
 where $\| \cdot \|_2$ and $\| \cdot \|_{\infty}$ denote matrix operation norm and largest element in magnitude. 
\end{prop}
The first term in (\ref{eq_11}) controls the rank of $\bm{\Theta}$ given a fixed sparsity level, where a larger value indicates a lower rank. The second term controls the sparsity level of $\bm{\Psi}$ given a fixed rank, where larger value indicates a lower sparsity level. Intuitively, networks $\bm{\Theta}$ and $\bm{\Psi}$ can be globally identified when they are well-separated in terms of either rank or sparsity.

\vspace{-5mm}
\subsection{Model estimation}

The model parameters $\Omega$ can be estimated via the constrained likelihood maximization as 
\begin{align*}
    \argmax_{\bm{\Omega} =  \{\bm{\Theta},\bm{\Psi},\bm{\pi}\}} 
    \frac{1}{C} \sum_{c=1}^C \log \mathbb{P}(\bm{t}^{(c)}; \bm{\Omega}) \quad \text{s.t. } & \bm{\Theta}\odot (\bm{I}-\bm{A}) = \bm{0}, \; \text{rank}(\bm{\Psi}) \leq r,
\end{align*}
where $\bm{I}$ is $N$-by-$N$ matrix with elements being $1$. 
However, both the likelihood function and rank regularization are difficult to directly optimize. Therefore, we maximize the evidence lower bound of the log-likelihood function and replace the low-rank penalty by its convex relaxation of nuclear norm $\|\cdot\|_{\star}$. The optimization problem can be reformulated as follows:
\begin{align*}
    \argmax_{\bm{q}(\bm{Z}), \bm{\Omega}} & \mathbb{E}_{\bm{q}(\bm{Z})} \big[\frac{1}{C}\sum_{c=1}^C \log \mathbb{P}(\bm{t}^{(c)},\bm{Z};\bm{\Omega})\big] - \big[ \mathbb{E}_{\bm{q}(\bm{Z})} \log \bm{q}(\bm{Z}) \big]\; 
    \text{s.t. } \bm{\Theta}\odot (\bm{I}-\bm{A}) = \bm{0}, \; \|\bm{\Psi}\|_{\star} \leq \rho,
\end{align*}
where $\mathbb{E}_{\bm{q}(\bm{Z})}$ is expectation of $\bm{Z}$ in terms of a factorizable distribution as $\bm{q}(\bm{Z}) =  \prod_{c=1}^C\prod_{i = 1}^N \mathbb{P}(Z_i^c \mid \bm{t}^{(c)}; \bm{\Omega})$. 
The above objective function can be optimized via EM-type algorithm by iteratively updating $\bm{q}(\bm{Z})$ and $\bm{\Omega}$. Specifically, given the parameters $\bm{\Omega}^{(m)}$ at the $m$-th step:
\begin{align*} 
    &\textbf{E-step:} \hspace{35mm} \bm{q}(\bm{Z};\bm{\Omega}^{(m)}) = \prod_{c=1}^C\prod_{i = 1}^N \mathbb{P}(Z_i^c \mid \bm{t}^{(c)}; \bm{\Omega}^{(m)})\\
    &\textbf{M-step:} \; \bm{\Omega}^{(m+1)} \! = \!\argmax_{\bm{\Omega}} \frac{1}{C}\sum_{c=1}^{C }\mathbb{E}_{\bm{q}(\bm{Z};\bm{\Omega}^{(m)})} \big[ \log  \mathbb{P}(\bm{t}^{(c)}\!,\bm{Z};\bm{\Omega})  \big]\;  
     \text{s.t. }  \bm{\Theta}\!\odot\! (\bm{I}-\bm{A}) = 0, \|\bm{\Psi}\|_{\star}\! \leq \!\rho.
\end{align*}
In E-step, the posterior distribution of network indicators $\hat{\pi}_i^{c} = \mathbb{P}(Z_i^c \mid \bm{t}^{(c)}; \bm{\Omega}^{(m)})$ can be explicitly updated via ($\ref{E_step}$).
In M-step, the objective function $\bm{Q}(\bm{\Omega}\mid \bm{\Omega}^{(m)}) := \frac{1}{C}\sum_{c=1}^{C }\mathbb{E}_{\bm{q}(\bm{Z};\bm{\Omega}^{(m)})} \big[ \log  \mathbb{P}(\bm{t}^{(c)},\bm{Z};\bm{\Omega})\big] $ can be decomposed as $\bm{Q}(\bm{\Omega}\mid \bm{\Omega}^{(m)}) = \bm{Q}_1(\bm{\Theta}\mid \bm{\Omega}^{(m)}) + \bm{Q}_2(\bm{\Psi}\mid \bm{\Omega}^{(m)}) + \bm{Q}_3(\bm{\pi}\mid \bm{\Omega}^{(m)})$, and $\bm{Q}_1(\bm{\Theta} \mid \bm{\Omega}^{(m)}) = \frac{1}{C}\sum_{c=1}^C \sum_{i=1}^N g_1(i,c)$, $\bm{Q}_2(\bm{\Psi} \mid \bm{\Omega}^{(m)}) = \frac{1}{C}\sum_{c=1}^C \sum_{i=1}^N g_2(i,c)$, $\bm{Q}_3(\bm{\pi} \mid \bm{\Omega}^{(m)}) = \frac{1}{C}\sum_{c=1}^C \sum_{i=1}^N g_3(i,c)$, where 
\begin{align*}
    &g_1(i,c) := \hat{\pi}_{i}^c\log\big\{ \mathbbm{1}( t^c_i<T) \mathbb{P}_{I}(t^c_i; \bm{\Theta}_{\cdot i}) + \mathbbm{1}( t^c_i\geq T) \mathbb{P}_{U}(T; \bm{\Theta}_{\cdot i})\big\}\\
    &g_2(i,c) := (1-\hat{\pi}_{i}^c)\log\big\{ \mathbbm{1}( t^c_i<T) \mathbb{P}_{I}(t^c_i; \bm{\Psi}_{\cdot i}) + \mathbbm{1}( t^c_i\geq T) \mathbb{P}_{U}(T; \bm{\Psi}_{\cdot i})\big\}\\
    &g_3(i,c): = \hat{\pi}_i^c\log \pi_i + (1 - \hat{\pi}_i^c)\log (1 - \pi_i),   
\end{align*}
and $\{\hat{\pi}_i^c\}$ depends on $\bm{\Omega}^{(m)}$ via (\ref{E_step}). Then $\bm{\Theta}$, $\bm{\Psi}$, and $\bm{\pi}$ can be updated parallelly as
\begin{align}
    &\text{M.1}: \bm{\Theta}^{(m+1)} =  \argmax_{\bm{\Theta}} \bm{Q}_1(\bm{\Theta}\mid \bm{\Omega}^{(m)})\; \text{ s.t. }\; \bm{\Theta}\odot (\bm{I}-\bm{A}) = \bm{0} \label{eq_12} \\
    &\text{M.2}: \bm{\Psi}^{(m+1)} =  \argmax_{\bm{\Psi}} \bm{Q}_2(\bm{\Psi}\mid \bm{\Omega}^{(m)})\; \text{ s.t. }\; \|\bm{\Psi}\|_{\star} \leq \rho \label{eq_12_2}\\
    &\text{M.3}: \bm{\pi}^{(m+1)} =  \argmax_{\bm{\pi}} \bm{Q}_3(\bm{\pi}\mid \bm{\Omega}^{(m)}) \label{eq_12_3}.
\end{align}
When the structural network $\bm{A}$ is not observed, one can impose $l_1$ norm penalty to pursue sparsity structure in $\bm{\Theta}$. Accordingly, the constraint in optimization (\ref{eq_12}) is replaced by $\| \bm{\Theta} \|_1 \leq s' $ where $s'>0$ is the sparsity tuning parameter. 
The main advantage of the proposed estimation is that the M-step becomes convex optimization problems. Specifically, denote parameter spaces $\bm{\Xi}_{\bm{\Theta}}(s): = \{ \bm{\Theta} \in  [0,\beta_1]^{N\times N} \mid \bm{\Theta}\odot (\bm{I}- \bm{A}) =\bm{0}\} $ with $s = \| \bm{A} \|_0$, $\bm{\Xi}_{\bm{\Psi}}(\rho):= \{  \bm{\Psi} \in  [0,\beta_2]^{N\times N} \mid  \| \bm{\Psi} \|_{\star} \leq \rho \}$, and $\bm{\Xi}_{\bm{\pi}}: = \{ \bm{\pi} \in [\epsilon,1-\epsilon]^N\}$ where $\beta_1,\beta_2$ are nonnegative constants and $0< \epsilon < 0.5$. We have the following result:  
\begin{theorem}
The parameter spaces $\bm{\Xi}_{\bm{\Theta}}(s)$, $\bm{\Xi}_{\bm{\Psi}}(\rho)$, and $\bm{\Xi}_{\bm{\pi}}$ are convex set for any $s>0,\; \rho>0$, and $\bm{Q}_3(\bm{\pi}\mid \bm{\Omega}^{(s)})$ is concave on $\bm{\pi}$. 
If the survival function $S$ is log-concave and hazard function $H$ is concave, 
then $\bm{Q}_1(\bm{\Theta}\mid \bm{\Omega}^{(s)})$ and $\bm{Q}_2(\bm{\Psi}\mid \bm{\Omega}^{(s)})$ are concave on $\bm{\Theta}$ and $\bm{\Psi}$, respectively. Furthermore, if for any node $i$, the probabilities of being source node $\bm{P}(v)>0$ for $v \in \mathcal{R}$ where $\mathcal{R}$ denotes the set of nodes from which $i$ is reachable via a directed path, then $\mathbb{E}_{\bm{t}}[\bm{Q}_1(\bm{\Theta}\mid \bm{\Omega}^{(s)})]$ and $\mathbb{E}_{\bm{t}}[\bm{Q}_2(\bm{\Psi}\mid \bm{\Omega}^{(s)})]$ are also strictly concave in terms of $\bm{\Theta}$ and $\bm{\Psi}$, respectively.  
\end{theorem}
The assumption on survival function can be satisfied by popular risk models including Exp, Pow, and Ray model, and other additive risk models. 
Theorem 3.1 guarantees that each sub-optimization (\ref{eq_12}), (\ref{eq_12_2}), and (\ref{eq_12_3}) in the M-step is convex optimization, and hence has an unique and optimal solution. The tuning parameter of low-rank penalty in (\ref{eq_12_2}) can be adaptively selected. Specifically, we first randomly separate the total cascade samples into a training subset $C_{train}$ and a validation subset $C_{val}$, and estimate model parameters $\bm{\hat{\Omega}}_{\mu}$ on $C_{train}$ given a specific $\mu$. Then we calculate the log-likelihood of validation samples as $\frac{1}{|C_{val}|}\sum_{c\in C_{val}} \log \mathbb{P}(\bm{t}^{(c)}, \bm{\hat{\Omega}}_{\mu})$, and select $\mu$ such that 
$$  \mu = \argmax_{\mu \in \bm{G}} \frac{1}{|C_{val}|}\sum_{c\in C_{val}} \log \bm{P}(\bm{t}^{(c)}, \bm{\hat{\Omega}}_{\mu}),$$
where $\bm{G}$ is a grid of candidates for penalty values. We summarize the optimization and more optimization details in Appendix.

\vspace{-5mm}
\section{Theory}

In this section, we establish both computational and statistical convergence of the model parameter estimation based on the concavity of auxiliary function $\bm{Q}(\bm{\Omega}\mid \bm{\Omega}')$. Let $\bm{\Omega}^{\star}: = \{\bm{\Theta}^{\star}, \bm{\Psi}^{\star}, \bm{\pi}^{\star} \} \in  \bm{\Xi}_{\Omega}(s,\rho)$ be the true model parameters where $\bm{\Xi}_{\Omega}(s,\rho):= \bm{\Xi}_{\bm{\Theta}}(s) \times \bm{\Xi}_{\bm{\Psi}}(\rho) \times  \bm{\Xi}_{\bm{\pi}}$. We follow the notations in Section \ref{sec-3}, and introduce parameter space metric as $\| \bm{\Omega} - \bm{\Omega}'\|^2: = \| \bm{\Theta} - \bm{\Theta}'\|^2_{F} + \| \bm{\Psi} - \bm{\Psi}'\|^2_{F} + \| \bm{\pi} - \bm{\pi}'\|^2_{2}$ where $\bm{\Omega}, \; \bm{\Omega}' \in \bm{\Xi}_{\Omega}$. Denote $\bm{\hat{\pi}}_{\bm{c}} = ( \hat{\pi}^c_1,\cdots, \hat{\pi}^c_N)$ where $\hat{\pi}^c_i$ is the posterior probability defined in (\ref{E_step}). To establish theoretic guarantee, assumptions required as as follows. 
  
\begin{assumption}
For any fixed $\bm{\Omega}, \; \bm{\Omega}' \in \bm{\Xi}_{\Omega}(s,\rho)$, there exists $\epsilon_{\bm{\Theta}}(C,\delta)>0,\epsilon_{\bm{\Psi}}(C,\delta)>0, \epsilon_{\bm{\pi}}(C,\delta)>0$ such that with probability at least $1 - \delta$: 
  $\|  \frac{\partial \bm{Q}_1(\bm{\Theta}\mid \bm{\Omega}')}{\partial \bm{\Theta}} - \mathbb{E}_{\bm{t}}(\frac{\partial \bm{Q}_1(\bm{\Theta}\mid \bm{\Omega}')}{\partial \bm{\Theta}} )\|_{F} \leq \epsilon_{\bm{\Theta}}(C,\delta)$,
$\|\frac{\partial \bm{Q}_2(\bm{\Psi}\mid \bm{\Omega}')}{\partial \bm{\Psi}} - \mathbb{E}_{\bm{t}}(\frac{\partial \bm{Q}_2(\bm{\Psi}\mid \bm{\Omega}')}{\partial \bm{\Psi}} )\|_{F} \leq \epsilon_{\bm{\Psi}}(C,\delta)$, 
$ \| \frac{1}{C}\sum_{c=1}^C \bm{\hat{\pi}}_{\bm{c}} - \mathbb{E}_{\bm{t}}(\bm{\hat{\pi}}_{\bm{c}}) \|_{2} \leq \epsilon_{\bm{\pi}}(C,\delta)$,
where $C$ is the number of cascade samples and $0<\delta<1$.
\end{assumption}
Assumption 1 quantifies the deviations of sample gradients of auxiliary function on $(\bm{\Theta},\bm{\Psi}, \bm{\pi})$ from their population counterparts, where $\epsilon_{\bm{\Theta}}, \epsilon_{\bm{\Psi}},\epsilon_{\bm{\pi}}$ characterize finite-samples statistical errors. Conditions similar to Assumption 1 are utilized in \citep{10.1214/16-AOS1435, yi2015regularized} to analyze convergence of EM algorithm.  

\begin{assumption}
  For any $\bm{\Omega}, \; \bm{\Omega}' \in \bm{\Xi}_{\Omega}$, there exists constant $\kappa_{\bm{\pi}}>0$ such that $\| \mathbb{E}_{\bm{t}}\big(\bm{\hat{\pi}_{c}}(\bm{\Omega})\big) - \mathbb{E}_{\bm{t}}\big(\bm{\hat{\pi}_{c}}(\bm{\Omega}')\big) \|_2^2 \leq  \kappa_{\bm{\pi}} \|  \bm{\Omega} - \bm{\Omega}'\|^2$. 
\end{assumption}
Assumption 2 states that the posterior probability of network indicator ${Z_i^c}$ is continuous over $\bm{\Omega}$. Intuitively, Assumption 2 plays the similar role of gradient smoothness condition in \citep{10.1214/16-AOS1435, yi2015regularized, wang2014high} to ensure that auxiliary function $ \mathbb{E}_{\bm{t}}(\frac{\partial\bm{Q}(\bm{\Omega\mid \bm{\Omega}'})}{\partial \bm{\Omega}})$ is smooth over $\bm{\Omega}'$. Notice that $\kappa_{\bm{\pi}}$ implicitly depends on the separability of true diffusion networks as $\| \bm{\Theta}^{\star} - \bm{\Psi}^{\star} \|_2$. 


\begin{assumption}
    The log-likelihoods of cascade $\bm{t}^{(c)}$ over networks $\bm{\Theta}$ and $\bm{\Psi}$ are $g_1(c): = \sum_{i=1}^Ng_1(i,c)$ and $g_2(c): = \sum_{i=1}^Ng_2(i,c)$. Let $\lambda_{min}$ denote the smallest eigenvalue.
    For all cascade samples $\{\bm{t}^{(c)}\}_{c=1}^C$,
    there exists constant $L>0$ such that 
    \begin{align*}
        \big| \lambda_{min}\Big( \frac{\partial^2 g_1(c)}{\partial\bm{\Theta}^2}   \Big) \big| \leq L, \; \big| \lambda_{min}\Big( \frac{\partial^2 g_2(c)}{\partial\bm{\Psi}^2} \Big) \big| \leq L. 
         \end{align*}
\end{assumption}
Assumption 3 is a regularity condition requiring eigenvalues are bounded, which guarantee the smoothness of $\bm{Q}(\bm{\Omega}\mid \bm{\Omega}')$ over $\bm{\Omega}$. Denote the smallest and largest eigenvalues of Hessian matrices $\mathbb{E}_{\bm{t}}\big[\frac{\partial^2 \bm{Q}_1(\bm{\Theta}\mid \bm{\Omega}')}{\partial\bm{\Theta}^2} \big]$ and $\mathbb{E}_{\bm{t}}\big[\frac{\partial^2 \bm{Q}_1(\bm{\Psi}\mid \bm{\Omega}')}{\partial\bm{\Psi}^2} \big]$ as $\lambda_{min}(\bm{\Theta},\bm{\Omega}'), \lambda_{min}(\bm{\Psi},\bm{\Omega}')$ and $\lambda_{max}(\bm{\Theta},\bm{\Omega}'), \lambda_{max}(\bm{\Psi},\bm{\Omega}')$. Notice that both smallest and largest eigenvalues are continuous functions of the corresponding Hessian matrices \citep{horn2012matrix}, which are continuous on $\bm{\Theta},\bm{\Psi},\bm{\Omega}'$.
Then there exist constants $0< \gamma_l < \gamma_u $ such that $|\max_{\bm{\Xi_{\Theta}}(s)\times \bm{\Xi_{\Omega}}}\lambda_{max}(\bm{\Theta},\bm{\Omega}')| \geq \gamma_l$, $|\max_{\bm{\Xi_{\Psi}}(\rho)\times \bm{\Xi_{\Omega}}}\lambda_{max}(\bm{\Psi},\bm{\Omega}')| \geq \gamma_l$, and $|\min_{\bm{\Xi_{\Theta}}(s)\times \bm{\Xi_{\Omega}}}\lambda_{min}(\bm{\Theta},\bm{\Omega}')| \leq \gamma_u$,  $|\min_{\bm{\Xi_{\Psi}}(\rho)\times \bm{\Xi_{\Omega}}}\lambda_{min}(\bm{\Psi},\bm{\Omega}')| \leq \gamma_u$ based on the strong concavity of $\mathbb{E}_{\bm{t}}(\bm{Q}_1(\bm{\Theta}\mid \bm{\Omega}'))$ and $\mathbb{E}_{\bm{t}}(\bm{Q}_2(\bm{\Psi}\mid \bm{\Omega}'))$ from Theorem 3.1, and compactness of $\bm{\Xi}_{\bm{\Theta}}(s) \times \bm{\Xi}_{\bm{\Psi}}(\rho) \times  \bm{\Xi}_{\bm{\pi}}$. 

The theoretic results depend on several quantities defined as follows. Due to the self-consistency property $\bm{\Omega}^{\star} = \argmax_{\bm{\Omega}\in \bm{\Xi}_{\Omega}(s,\rho)} \mathbb{E}_{\bm{t}}\bm{Q}(\bm{\Omega} \mid \bm{\Omega}^{\star})$, we denote the oracle estimation as $\bm{\hat{\Omega}}: = \argmax_{\bm{\Omega}\in \bm{\Xi}_{\Omega}(s,\rho)} \bm{Q}(\bm{\Omega}\mid \bm{\Omega}^{\star})$ when the posterior distributions of network indicators $\bm{Z}$ are known. 
The results also depend on the topology of diffusion networks. Let $\bm{B} = \{(i,j)\mid \bm{\Psi}^{\star}_{ij}>0\}$ be the support of latent network $\bm{\Psi}^{\star}$. We introduce $d_1  = \max\{ \max_{i}\{\bm{B}_{i\cdot}\}, \max_{i}\{\bm{B}_{\cdot i}\}\}$ and $d_2 = \max\{ \max_{i,j}\{ \langle \bm{B}_{i\cdot}, \bm{B}_{j\cdot} \rangle \}, \max_{i,j}\{ \langle \bm{B}_{\cdot i}, \bm{B}_{\cdot j} \rangle  \}\}$. Intuitively, $d_1$ indicates the largest node in-degree and out-degree on the latent network, and $d_2$ denotes the largest number of shared direct parents or direct children between any two nodes over the latent network.
\begin{theorem}
    Denote $s^{\star} = \| \bm{\Theta}^{\star}\|_0,\; \rho^{\star} = \| \bm{\Psi}^{\star} \|_{\star}$ and $r := \text{rank}(\Psi)$.
    Under Assumption 1-3 and assumptions in Theorem 3.1, and assume that 1) $\kappa_{\bm{\pi}}\leq \kappa_0$ for a constant $\kappa_0(\gamma_l,\gamma_u,\eta)\leq \min\{ \sqrt{\gamma_l},\frac{4\gamma_u^2}{\eta^2}\}$, 2) the number of cascade samples $C = \omega \big(\log (\frac{N^2m}{\delta})^{L/\gamma_l} \big)$ and 3) $\rho \leq \rho^{\star}$, then there exists a constant $0< \kappa(\kappa_{\bm{\pi}}, \gamma_l,\gamma_u,\eta) < 1$ such that 
    for the  $m$-th iteration estimation $\bm{\Omega}^{(m)}$ from EM Algorithm, we have with probability at least $1 - 15\delta$,
    \begin{align} \label{theo_1}
     \|{\bm{\Omega}}^{(m)} - {\bm{\Omega}}^{\star} \|^2 &\leq 
    2\kappa^m \|{\bm{\Omega}}^{(0)} - \hat{\bm{\Omega}}\|^2  + \Delta_{optim}  + \big(2+ \frac{4}{9}\Delta + \frac{12\kappa_{\bm{\pi}}}{1-\kappa}\big)(\Delta^{\Theta}_{est} + \Delta^{\Psi}_{est} + \Delta^{\pi}_{est})\\
\Delta_{optim}  & =  \frac{\Delta}{\eta^2{\kappa_{\bm{\pi}}}}(\epsilon^2_{\bm{\Theta}}(C,\delta/m)+\epsilon^2_{\bm{\Psi}}(C,\delta/m)) +  \frac{12}{1-\kappa}\epsilon^2_{\bm{\pi}}(C,\delta/m), \nonumber \\
\Delta^{\Theta}_{est}   = & \frac{c_1 s^{\star}  \log (s^{\star}/\delta)}{\gamma^2_l{C}}, \; \Delta^{\Psi}_{est} = \frac{c_2r\sqrt{N^2d_2+Nd_1} \log (2N/\delta)}{\gamma^2_l{C}},\;   \Delta^{\pi}_{est} = \frac{c_3N \log (N/\delta)}{\epsilon^2(1-\epsilon)^2{C}}, \nonumber
\end{align}
where $\Delta = \frac{36\eta \sqrt{\kappa_{\bm{\pi}}}}{1-\kappa}(3\gamma_u - \eta\sqrt{\kappa_{\bm{\pi}}})^{-1}$, $\eta: = \max\{ \sqrt{\max_{i} \mathbb{E}_{\bm{t}}(\| \frac{\partial g_1(i,c)}{\partial \bm{\Theta}_{\cdot i} } \|_2^2)  }, \sqrt{\max_{i} \mathbb{E}_{\bm{t}}(\| \frac{\partial g_2(i,c)}{\partial \bm{\Psi}_{\cdot i} } \|_2^2)  } \}$
and $c_1,c_2,c_3>0$ are constants. And $A = \omega(B)$ means $A\geq cB$ for some constant $c>0$. 
\end{theorem}
Theorem 4.1 indicates that the convergence of EM estimation to true parameters consists of three terms on the RHS of (\ref{theo_1}). The first term refers to the computational error, which decays at a geometric rate if the $\kappa_{\bm{\pi}}$ is small enough, i.e., the auxiliary function $\bm{Q(\bm{\Omega\mid \bm{\Omega}}')}$ is smooth enough over the second argument. The result is consistent with optimization error established in \citep{wang2014high, 10.1214/16-AOS1435}. Both the second term and third term are contributed by the statistical error originated from the cascading sample randomness. Specifically, $\Delta_{optim}$ characterizes the deviation of the sample version of auxiliary function $\bm{Q}(\bm{\Omega}\mid \bm{\Omega}')$ from its population
version $\mathbb{E}_{\bm{t}}[\bm{Q}(\bm{\Omega}\mid \bm{\Omega}')]$ in the optimization. The term $\Delta^{\Theta}_{est}  + \Delta^{\Psi}_{est}  + \Delta^{\pi}_{est} $ quantifies the estimation error of the oracle estimator $\bm{\hat{\Omega}}$.  Compared with existing convergence analysis of EM algorithm \citep{wang2014high, 10.1214/16-AOS1435}, Theorem 4.1 does not require that initial $\bm{\Omega}^{(0)}$ is close to $\bm{\Omega}^{\star}$ by taking the advantage of the global concavity of the auxiliary function.  

\textit{Remark 1}: Compared with results in \citep{wang2014high, 10.1214/16-AOS1435}, Theorem 4.1 provides a more refined analysis by  decomposing statistical error into optimization term and estimation term, 
and concretizes the estimation error term. The estimation error consists of three terms $\frac{s^{\star}\log(s^{\star})}{{C}}$, $\frac{r \sqrt{N^2d_2 + Nd_1}\log N }{C}$, and $\frac{N\log N}{C}$, corresponding to the convergence rates of estimating $\bm{{\Theta}}$, $\bm{{\Psi}}$, and $\bm{\pi}$, respectively. Specifically, the convergence rate of $\bm{\hat{\Theta}}$ is consistent with the Frobenius norm bounds for graphical Lasso \citep{10.1214/08-EJS176, wainwright2019high} based on the fact that $s^{\star} = \mathcal{O}(N^2)$. The convergence rate of $\bm{\hat{\Psi}}$ depends on the topological structure of $\bm{\Psi}^{\star}$, which is dominated by $d_2$. Intuitively, larger $d_2$ indicates a higher overlap between two nodes in terms of parent and child nodes. Therefore, nodes tend to have similar activation patterns across cascade samples, and it becomes more difficult to estimate transmission rates of nodes' own parent and child nodes. 
When $d_2$ is bounded then the rate of $\bm{\hat{\Psi}}$ simplifies to $\frac{rN\log N}{C}$, which is consistent with the optimal rate of low-rank estimation and matrix completion under Frobenius norm \citep{10.1214/10-AOS860, negahban2012restricted}.      

\textit{Remark 2}: The length of observation window $T$ implicitly affects the convergence rate via $\gamma_l$. Notice that $\frac{\partial^2 \log S(T\mid t_j;\mathcal{E}_{ji})}{\partial\mathcal{E}^2} = 0$ for the survival models, then the values of 
$\frac{\partial^2 \bm{Q}_1(\bm{\Theta}\mid \bm{\Omega}')}{\partial\bm{\Theta}^2}$ and $\frac{\partial^2 \bm{Q}_2(\bm{\Psi}\mid \bm{\Omega}')}{\partial\bm{\Psi}^2}$ are determined by the activated nodes. As $T$ decreases, the number of activated nodes decreases, and hence the Frobenius norms of $\mathbb{E}_{\bm{t}}\big[\frac{\partial^2 \bm{Q}_1(\bm{\Theta}\mid \bm{\Omega}')}{\partial\bm{\Theta}^2} \big]$ and $\mathbb{E}_{\bm{t}}\big[\frac{\partial^2 \bm{Q}_2(\bm{\Psi}\mid \bm{\Omega}')}{\partial\bm{\Psi}^2} \big]$ decrease to zero. Then the largest eigenvalue $\gamma_l$ decreases to zero, which slows the convergence rate.


The convergence of EM algorithm relies on the strong concavity of sample version of auxiliary $\bm{Q}_1(\bm{\Theta}\mid \bm{\Omega})$ and $\bm{Q}_2(\bm{\Psi}\mid \bm{\Omega})$ over $\bm{\Theta}$ and $\bm{\Psi}$, which might be difficult to satisfy
when the sample size is not large enough. On the other hand, the support constraint on $\bm{\Theta}$ and low-rank constraint on $\bm{\Psi}$ in (\ref{eq_12}) can limit the feasible solution into a lower dimensional subspace on which the sample auxiliary functions remain strictly concave.
\begin{assumption}
    For any $\bm{\Omega}, \; \bm{\Omega}' \in \bm{\Xi}_{\Omega}$, let $L_1(\bm{\Theta};\bm{\Theta}') = \bm{Q}_1(\bm{\Theta}\mid \bm{\hat{\Omega}}) - \bm{Q}_1(\bm{\Theta}'\mid \bm{\hat{\Omega}}) - \langle   \nabla \bm{Q}_1(\bm{\Theta}' \mid \bm{\hat{\Omega}}),  \bm{\Theta} - \bm{\Theta}' \rangle$ and $L_2(\bm{\Psi};\bm{\Psi}') = \bm{Q}_1(\bm{\Psi}\mid \bm{\hat{\Omega}}) - \bm{Q}_1(\bm{\Psi}'\mid \bm{\hat{\Omega}}) - \langle   \nabla \bm{Q}_2(\bm{\Psi}' \mid \bm{\hat{\Omega}}),  \bm{\Psi} - \bm{\Psi}' \rangle$. There exists $0<\gamma_l\leq \gamma_u, \tau_l>0,\tau_u>0$ with probability at least $1-\delta$ such that
    \begin{align*}
        - \gamma_u \| \bm{\Theta} - \bm{\Theta}' \|_F^2 \leq &L_1(\bm{\Theta};\bm{\Theta}') \leq  - \gamma_l \| \bm{\Theta} - \bm{\Theta}' \|_F^2, \\
        - \tau_{u}\| \bm{\Psi} - \bm{\Psi}'   \|^2_{\star} - \gamma_u \| \bm{\Psi} - \bm{\Psi}' \|_F^2 \leq &L_2(\bm{\Psi};\bm{\Psi}') \leq  - \gamma_l \| \bm{\Psi} - \bm{\Psi}' \|_F^2 + \tau_{l}\| \bm{\Psi} - \bm{\Psi}'   \|^2_{\star}
    \end{align*}
\end{assumption}
Assumption 4 refers to the restricted strong concavity and restricted smooth conditions \citep{10.1214/12-STS400, agarwal2010fast}, which are common in the constrained optimization analysis. Intuitively, Assumption 4 indicates the non-zero elements in $\bm{\Theta}$ are small enough such that the Hessian matrix of $\bm{Q}_1$ is well-conditioned. In terms of $\bm{\Psi}$, the Hessian matrix of $\bm{Q}_2$ becomes well-conditioned when adding the nuclear norm regularizer on $\bm{\Psi}$. 
\begin{corollary_1}
   Under Assumption 1,2,4 and assumptions in Theorem 3.1, and assume that 1) $\kappa_{\bm{\pi}}\leq \kappa_0$ for a constant $\kappa_0(\gamma_l,\gamma_u,\eta)\leq \min\{ \sqrt{\gamma_l},\frac{4\gamma_u^2}{\eta^2}\}$, 2) $\| \hat{\bm{\Psi}}\|_{\star} = \rho \leq \rho^{\star}$, and 3) $r\leq \frac{\gamma_l}{16\tau_l}$ then there exists a constant $0< \kappa(\kappa_{\bm{\pi}}, \gamma_l,\gamma_u,\eta, \phi) < 1$ such that 
    with probability at least $1 - 11\delta$
    \begin{align}\label{cor_1}
        \| \bm{\Omega}^{(m)} - \bm{\Omega}^{\star}\|^2  & \leq 2\kappa^m \| \bm{\Omega}^{(0)} - \bm{\hat{\Omega}}\|^2 + \Delta_{optim}  + C_0( \Delta^{\Theta}_{est} + \Delta^{\pi}_{est}) + (C_0+C_1)\Delta^{\Psi}_{est},\\
        \Delta_{optim}  & =  \frac{\Delta}{\eta^2{\kappa_{\bm{\pi}}}}(\epsilon^2_{\bm{\Theta}}(C,\delta/m)+ \phi\epsilon^2_{\bm{\Psi}}(C,\delta/m)) +  \frac{12}{1-\kappa}\epsilon^2_{\bm{\pi}}(C,\delta/m) \nonumber
\end{align}
where $C_0 = 2+ \frac{12\kappa_{\bm{\pi}}}{1-\kappa} + \frac{4}{9}\Delta(1+\phi)$, $C_1 = \frac{\phi\Delta(400r\tau_u+100r\tau_l)}{18\eta\sqrt{\kappa_{\bm{\pi}}}}$, $\phi := \frac{3\gamma_u - \eta\sqrt{\kappa_{\bm{\pi}}}}{3\gamma_u - \eta\sqrt{\kappa_{\bm{\pi}}} - 32r\tau_u}$, and $\Delta, \Delta^{\Theta}_{est}, \Delta^{\Psi}_{est}, \Delta^{\pi}_{est}$ are defined in Theorem 4.1. 
\end{corollary_1}
Corollary 4.2 relaxes the bounded eigenvalue requirement on the Hessian matrix of auxiliary function $\bm{Q}_2$ by the restricted strong concavity and smooth condition in Assumption 4, and therefore removes the requirement on the number of cascade samples $C$ in Theorem 4.1. At the cost, the convergence rate of $\bm{\hat{\Psi}}$ is slower than the rate in Theorem 4.1, and depends on the degrees of regularization $\tau_u$ and $\tau_l$.  

\noindent \textit{Remark 1}: We can further relax bounded eigenvalue assumption on the Hessian matrix of $\bm{Q}_1(\bm{\Theta\mid \bm{\Omega}'})$ as $-\tau_u \| \bm{\Theta} -  \bm{\Theta}'\|_1 - \gamma_u \| \bm{\Theta} - \bm{\Theta}' \|_F^2 \leq L_1(\bm{\Theta};\bm{\Theta}') \leq  - \gamma_l \| \bm{\Theta} - \bm{\Theta}' \|_F^2 + \tau_l \| \bm{\Theta} -  \bm{\Theta}'\|_1$, then $\bm{Q}_1$ becomes strong concavity and smooth after adding $l_1$-regularizer. Similar convergence rate can be obtained with the same order to the estimation errors in (\ref{cor_1}).  


\vspace{-5mm}

\section{Simulation}

In this section, we investigate the performance of the proposed method in recovering diffusion networks based on simulated cascade data. For a diffusion network $\mathcal{E}$, we measure the global transmission rate recovery accuracy of an estimation $\hat{\mathcal{E}}$ by normalized mean absolute error (MAE) as $\text{MAE}(\mathcal{{E}}) = \sum_{(i,j):\mathcal{E}_{ij}>0}\frac{|\mathcal{\hat{E}}_{ij} - \mathcal{E}_{ij}|}{\mathcal{E}_{ij}}$. In addition, we investigate the performance of topology recovery of the diffusion network, which is measured by three metrics including accuracy $\text{Acc}(\mathcal{E}) = \frac{\sum_{i,j}|\mathbb{I}(\mathcal{\hat{E}}_{ij}) - \mathbb{I}(\mathcal{E}_{ij})|}{\sum_{i,j}\mathbb{I}(\mathcal{\hat{E}}_{ij}) + \sum_{i,j} \mathbb{I}(\mathcal{E}_{ij})}$, precision $\text{Pre}(\mathcal{E}) = \frac{\sum_{i,j}\mathbb{I}(\mathcal{\hat{E}}_{ij}) \cdot \mathbb{I}(\mathcal{E}_{ij})}{\sum_{i,j} \mathbb{I}(\mathcal{\hat{E}_{ij}})}$, and recall $\text{Recall}(\mathcal{E}) = \frac{\sum_{i,j}\mathbb{I}(\mathcal{\hat{E}}_{ij}) \cdot \mathbb{I}(\mathcal{E}_{ij})}{\sum_{i,j} \mathbb{I}(\mathcal{E}_{ij})}$, where $\mathbb{I}(\alpha) = 1$ if $\alpha>0$ and $\mathbb{I}(\alpha) = 0$ otherwise. The estimation accuracy of diffusion network probabilities $\bm{\pi}$ are measured as $\text{MAE}(\bm{\pi}) = \sum_{i}\frac{|\hat{\pi}_{i} - \pi_{i}|}{\pi_{i}}$. In the following simulations, we initialize the diffusion networks as $\bm{\Theta^{(0)}} = \mathcal{E} \odot \bm{A}$ and $\bm{\Psi^{(0)}} = \mathcal{E} \odot (\bm{1} - \bm{A})$ where $\mathcal{E}$ is obtained from NetRate \citep{rodriguez2014uncovering}. 


\vspace{-5mm}
\subsection{Benchmarks comparison under different network topology}

In this subsection, we compare the proposed method with existing diffusion network recovery methods including NetRate \citep{rodriguez2014uncovering}, MMRate \citep{wang2014mmrate}, and ConNIe \citep{myers2010convexity}. ConNIe employs a maximum likelihood formulation via convex programming, incorporating an $l_1$-type penalty to promote sparsity in network estimation. Building upon ConNIe, NetRate explicitly represents diffusion as a continuous-time probabilistic process, characterized by edge-specific transmission rates governing diffusion probabilities. MMRate further extends this framework by accommodating multiple distinct diffusion patterns, assuming multiple heterogeneous diffusion networks. Both these benchmark methods are designed based on cascade generation processes following survival data modeling and independent cascade model. 

We generate diffusion networks with different topology to mimic real-world networks. Specifically, we consider the structural diffusion network $\bm{\Theta}$ to be random network, network with community structure, and scale-free network. 
With $\bm{A}$ being support of $\bm{\Theta}$, we generate $\bm{A}$ via Erdos-Renyi model with generation probability of edge being $0.01$ for random network $\bm{\Theta}$. For community structure, we generate $\bm{A}$ by stochastic block model with four equally-sized communities. The generation probability of within-community edge is $0.05$ and that of between-community edge is $0.01$. For scale-free network, we generate $\bm{A}$ via Barabási–Albert model where we set the number of edges to attach from a new node to existing nodes to be 1. Given the generated support $\bm{A}$, we construct $\bm{\Theta}$ by assigning each non-zero element of $\bm{A}$ a weight that follows uniform distribution between 1 to 5. We generate the latent diffusion network as $\bm{\Psi} = \Psi_1\Psi_2^{\top}$ where $\Psi_1, \Psi_2 \in \mathbb{R}_{+}^{N\times 5}$. The  proportion of non-zero elements in $\Psi_1, \Psi_2$ is set at $0.1$, and the non-zero elements are sampled from uniform distribution between 1 and 2. The latent diffusion network $\bm{\Psi}$ generated accordingly has an edge density of 0.05 with edge-wise transmission rates ranging from 1 to 8. In addition, we impute edges on $\bm{A}$ to ensure that each row and column of $\bm{\Theta} + \bm{\Psi}$ has at least one non-zero element to guarantee the connectedness of the combined network. We fix the size of diffusion networks at $N = 200$ and generate independent cascade samples based on Exp hazard model and $\bm{\Theta, \bm{\Psi}}$ with observation window length being $T=10$. 

\vspace{-2mm}
\begin{figure}[htbp]
\centering
\begin{subfigure}{0.5\textwidth}
  \centering
  \includegraphics[width=0.97\textwidth]{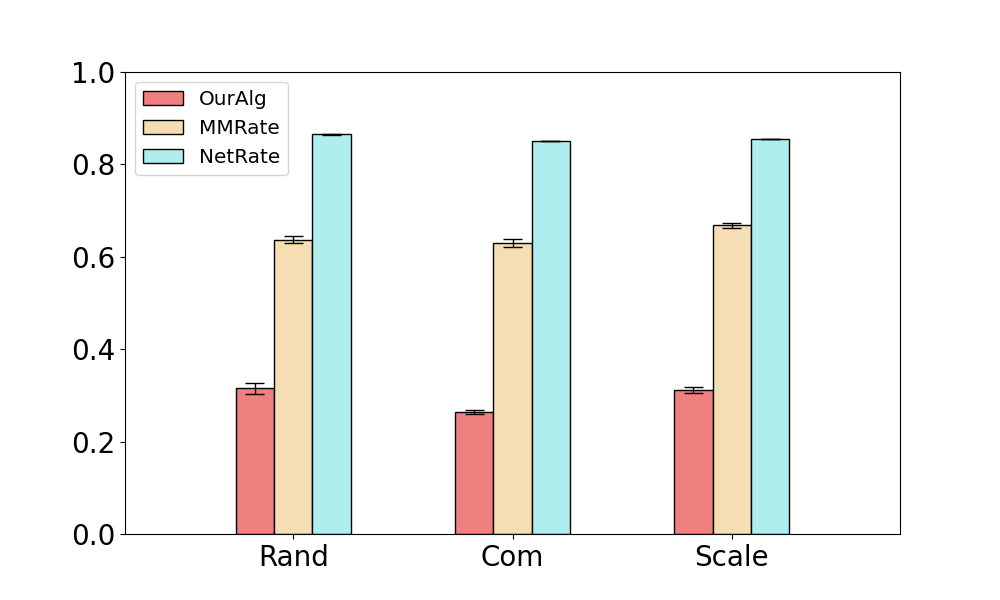}
  \vspace{-5mm}
  \caption{MAE$(\bm{\Theta})$}
\end{subfigure}%
\begin{subfigure}{0.5\textwidth}
  \centering
  \includegraphics[width=0.97\textwidth]{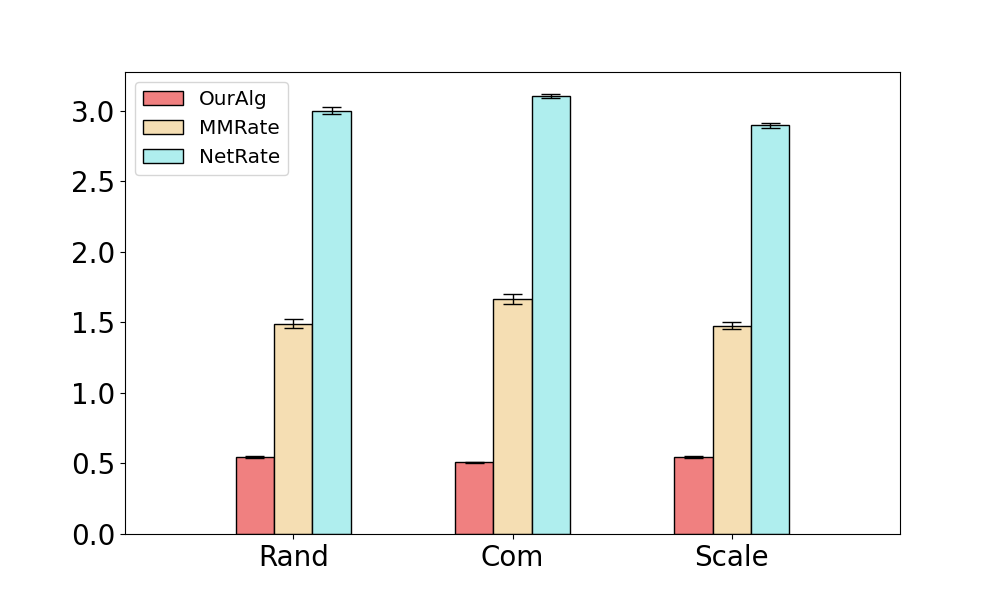}
  \vspace{-5mm}
  \caption{MAE$(\bm{\Psi})$}
\end{subfigure}%
\vspace{-5mm}
\caption{\small MAE of transmission rate estimations from different methods under three $\bm{\Theta}$ topologies. Rand: random; Com: community structure; Scale: scale-free.}
\label{benchmark:mae}
\end{figure}

\begin{figure}[htbp]
\centering
\begin{subfigure}{0.5\textwidth}
  \centering
  \includegraphics[width=0.97\textwidth,height = 4.3cm]{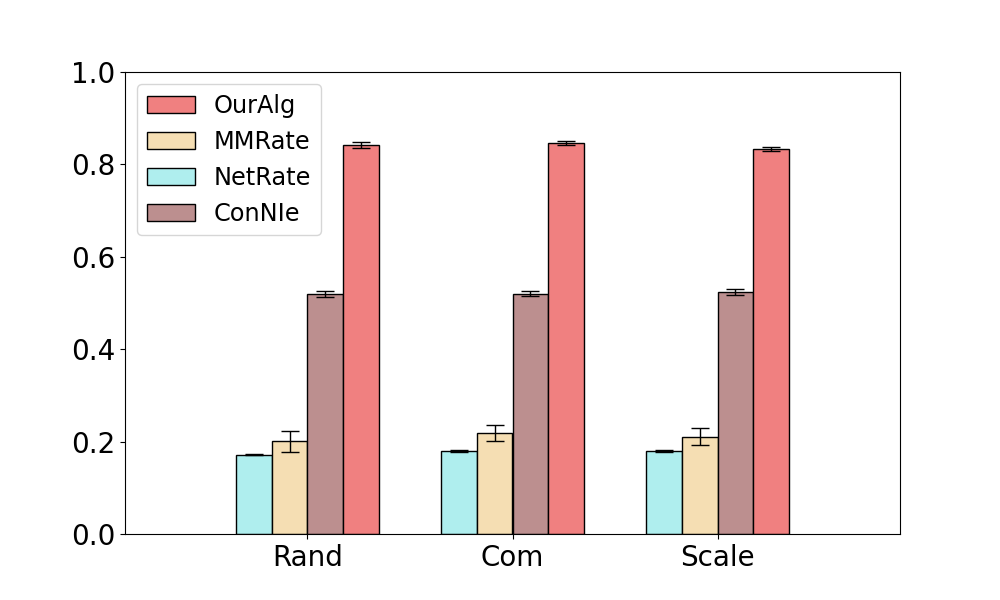}
  \vspace{-5mm}
  \caption{Accuracy$(\bm{\Psi})$}
\end{subfigure}%
\begin{subfigure}{0.5\textwidth}
  \centering
  \includegraphics[width=0.97\textwidth,height = 4.3cm]{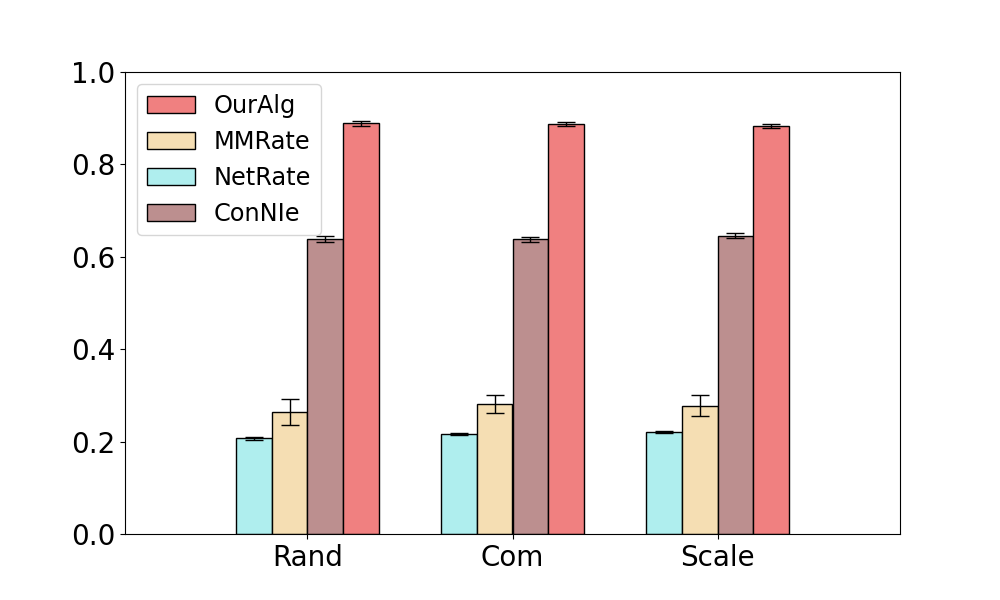}
  \vspace{-5mm}
  \caption{Accuracy$(\bm{\Theta} \cup \bm{\Psi})$}
\end{subfigure} \\
\vspace{-2mm}
\centering
\begin{subfigure}{0.5\textwidth}
  \centering
  \includegraphics[width=0.97\textwidth, height = 4.3cm]{BenchmarkTest/acc.png}
  \vspace{-5mm}
  \caption{Precision$(\bm{\Psi})$}
\end{subfigure}%
\begin{subfigure}{0.5\textwidth}
  \centering
  \includegraphics[width=0.97\textwidth, height = 4.3cm]{BenchmarkTest/acc_union.png}
  \vspace{-5mm}
  \caption{Precision$(\bm{\Theta} \cup \bm{\Psi})$}
\end{subfigure} \\
\vspace{-2mm}
\centering
\begin{subfigure}{0.5\textwidth}
  \centering
  \includegraphics[width=0.97\textwidth, height = 4.3cm]{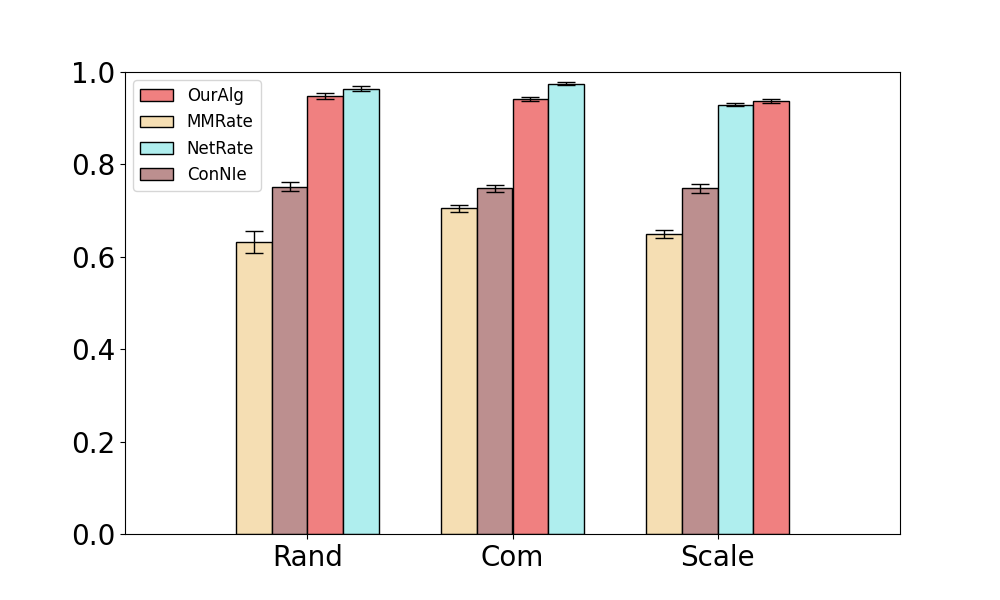}
  \vspace{-5mm}
  \caption{Recall$(\bm{\Psi})$}
\end{subfigure}%
\begin{subfigure}{0.5\textwidth}
  \centering
  \includegraphics[width=0.97\textwidth, height = 4.3cm]{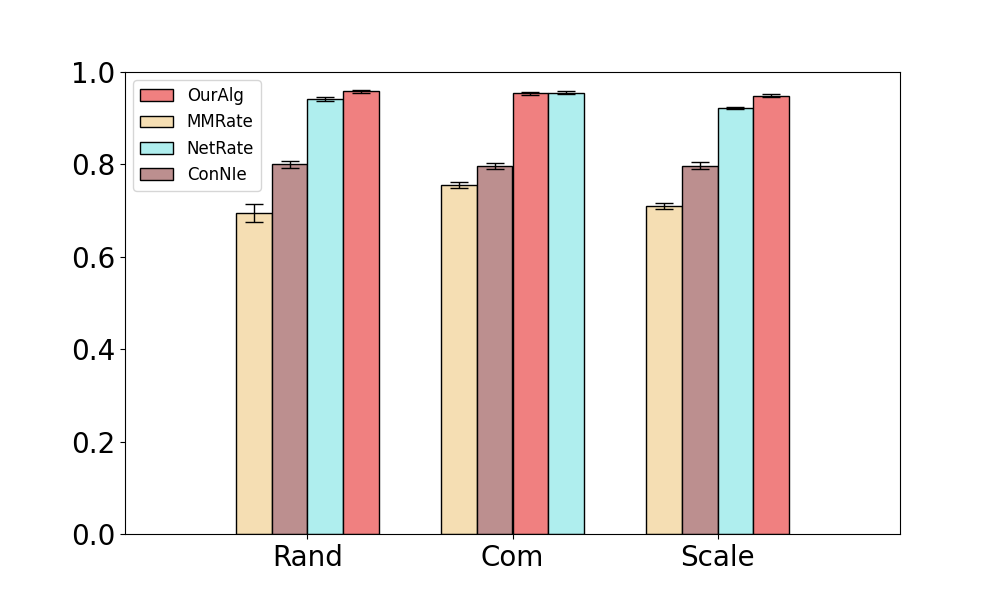}
  \vspace{-5mm}
  \caption{Recall$(\bm{\Theta} \cup \bm{\Psi})$}
\end{subfigure}
\vspace{-15mm}
\caption{\small Accuracy, precision, and recall of network topology estimation from different methods under different $\bm{\Theta}$ topologies.}
\label{benchmark:top}
\end{figure}

\vspace{-5mm}
Figure \ref{benchmark:mae} shows the MAE of transmission rate estimation on structural diffusion network $\bm{\Theta}$ under three network topologies, and corresponding MAE of latent diffusion network $\bm{\Psi}$. The proposed method (OurAlg) outperforms NetRate and MMRate by achieving lower $MAE$ on both $\bm{\Theta}$ and $\bm{\Psi}$. Notice that the MAE comparison does not include ConNIe since it does not estimate edge-wise transmission rates and only estimates the network topology. In addition, we compare the proposed method with benchmark methods in term of network topology recovery. 
Given that benchmark methods NetRate and ConNIe do not distinguish different diffusion networks by design, we investigate topology recovery of both the latent diffusion network $\bm{{\Psi}}$ and the joint network $\bm{{\Theta}}\cup \bm{{\Psi}}$ for fair comparison. 
Figure \ref{benchmark:top} illustrates the accuracy, precision, and recall from different methods, respectively. Under different network topologies, the proposed method achieves higher accuracy for recovering $\bm{{\Psi}}$ and joint network $\bm{{\Theta}} \cup \bm{{\Psi}}$ compared with NetRate, MMRate, and ConNIe. Specifically, our method significantly outperforms other methods in precision while remains similar or higher recall. 
We also investigate the estimation accuracy of network probability $\bm{\pi}$. Due to that only MMRate differentiate diffusion networks and estimate probabilities of cascading diffusing over a specific network, we compare $\text{MAE}(\bm{\pi})$ from the proposed method and MMRate. 
Table \ref{tab:pi} shows that our method outperforms MMRate in estimation of network probability under different network topologies. 
\begin{table}[htbp]
  \centering
  \scalebox{0.8}{
  \begin{tabular}{|c|c|c|c|}
    \hline
    $\text{MAE}(\bm{\pi})$ & Rand & Com & Scale \\
    \hline
    OurAlg & 0.047(0.002) & 0.049(0.002) & 0.050(0.002) \\
    \hline
    MMRate & 0.888(0.133) & 0.881(0.137) & 0.928(0.028) \\
    \hline
  \end{tabular}}
  \caption{\small MAE($\bm{\pi}$) of the proposed method and MMRate under three $\bm{\Theta}$ topologies (standard deviation in parenthesis).}
  \label{tab:pi}
\end{table}

\vspace{-5mm}

\subsection{Network recovery under different transmission models}

In this subsection, we investigate the performance of our diffusion network estimation when cascade samples are generated from popular transmission models including Exp, Pow, and Ray models. 
The diffusion networks $\bm{\Theta}$ and  $\bm{\Psi}$ are generated following the similar setup in section 5.1 with $\bm{\Theta}$ being random network and $\bm{\Psi}$ being low-rank. The detailed data generation procedures are provided in Appendix. 
The resulting $\bm{\Theta}$ and $\bm{\Psi}$ have edge density at 0.01 and 0.05, respectively. Based on $\bm{\Theta}$ and $\bm{\Psi}$, we generate different numbers of cascade samples at $C = 500, 1000, 1500, 2000$ with a fix observation window length at $T = 10$. In the Pow transmission model, we select delay parameter $\delta = 1$. 
\begin{table}[htbp]
\centering
\scalebox{0.75}{
\begin{tabular}{clcccccc}
\toprule
 &     &  {MAE$_{\bm{\Theta}}$} & {MAE$_{\bm{{\Psi}}}$} & MAE$_{\bm{{\pi}}}$ & {Acc$_{\bm{{\Psi}}}$} & {Pre$_{\bm{{\Psi}}}$} & {Rec$_{\bm{{\Psi}}}$}  \\
\midrule
\multirow{4}{*}{Exp}
  & C = 500 & 0.320(0.019) & 0.871(0.019) & 0.095(0.006) & 0.564(0.010) & 0.431(0.011) & 0.817(0.010)\\
  & C = 1000 & 0.235(0.013)& 0.667(0.007) & 0.091(0.005) & 0.743(0.009)& 0.653(0.014) & 0.862(0.007)\\
  & C = 1500 & 0.216(0.012) & 0.601(0.008) & 0.085(0.004) & 0.817(0.006) & 0.746(0.008) & 0.903(0.006)\\
  & C = 2000 & 0.187(0.011) & 0.562(0.007) & 0.082(0.004) & 0.857(0.004) & 0.794(0.006) & 0.930(0.005)\\
\cmidrule(lr){1-8}
\multirow{4}{*}{Ray}
  & C = 500 & 0.287(0.018) & 1.053(0.056) & 0.152(0.008) & 0.605(0.018) & 0.525(0.024) & 0.714(0.013)\\
  & C = 1000 & 0.211(0.015) & 0.990(0.100) & 0.130(0.008) & 0.712(0.019) & 0.605(0.026) & 0.866(0.012)\\
  & C = 1500 & 0.179(0.011) & 0.629(0.091) & 0.134(0.004) & 0.795(0.017) & 0.702(0.026) & 0.916(0.007)\\
  & C = 2000 & 0.188(0.012) & 0.605(0.081) & 0.137(0.004) & 0.807(0.011) & 0.709(0.019) & 0.937(0.011)\\
\cmidrule(lr){1-8}
\multirow{4}{*}{Pow}
  & C = 500 & 0.171(0.008) & 0.433(0.012) & 0.144(0.010) & 0.882(0.015) & 0.832(0.024) & 0.939(0.009)\\
  & C = 1000 & 0.129(0.006) & 0.328(0.007) & 0.131(0.007) & 0.956(0.004) & 0.932(0.007) & 0.982(0.004)\\
  & C = 1500 & 0.112(0.005) & 0.290(0.004) & 0.125(0.008) & 0.969(0.002) & 0.946(0.004) & 0.993(0.002)\\
  & C = 2000 & 0.106(0.006) & 0.271(0.004) & 0.129(0.011) & 0.973(0.003) & 0.950(0.005) & 0.997(0.001)\\
\bottomrule
\end{tabular}}
\caption{\small Diffusion network estimations from the proposed method under different cascade models and sample sizes.}
\label{tab:casmodel}
\end{table}
Table \ref{tab:casmodel} illustrates the transmission rates estimation and network topology recovery on $\bm{\Psi}$ under different transmission models and cascade sample sizes $C$. As the sample size $C$ increases, both the parameter estimations (MAE) and topology recovery metrics (Acc, Pre, Rec) improve under different transmission models. In addition, the degree of improvement decreases as more cascade samples become available. 
This pattern indicates the consistency of the proposed diffusion network estimators, and the convergence of the proposed EM-type optimization. Notice that the diffusion network recovery based on cascade samples generated from Pow transmission model is better than Exp and Ray model. This is because the Pow model imposes lower bound on activation time lag, which lowers the variation of activation time and introduces additional sparsity regularization.   

\vspace{-5mm}
\subsection{Network recovery under different support overlap}

In this subsection, we investigate the performance of the 
proposed estimation as the degree of overlapping between $\bm{\Theta}$ and $\bm{\Psi}$ varies, which is measured as $\text{overlap}(\bm{\Theta}, \bm{\Psi}) = \frac{\sum_{i,j}\mathbb{I}(\bm{{A}}_{ij}\times \bm{{B}}_{ij})}{\sum_{i,j}\mathbb{I}(\bm{{A}}_{ij} + \bm{{B}}_{ij})}$ where $\bm{A},\bm{B}$ are support of $\bm{\Theta},\bm{\Psi}$, respectively. We first generate latent diffusion network $\bm{\Psi}$ following a similar procedure in Section 5.1, and then generate $\bm{\Theta}$ so that $\text{overlap}(\bm{\Theta}, \bm{\Psi})$ has three levels at about $0.1,0.3,0.5$. Based on $\bm{\Theta}$ and $\bm{\Psi}$, we generate cascade samples from Exp transmission model with observation window length at $T=10$. The detailed generation procedure is provided in Appendix. 
\begin{table}[htbp]
\centering
\scalebox{0.75}{
\begin{tabular}{clcccccc}
\toprule
$\text{overlap}(\bm{\Theta}, \bm{\Psi})$  &    & \textbf{MAE$_{\bm{\Theta}}$} & \textbf{MAE$_{\bm{\Psi}}$} & \textbf{MAE$_{\bm{\pi}}$} & \textbf{Acc$_{\bm{\Psi}}$} & \textbf{Pre$_{\bm{\Psi}}$} & \textbf{Rec$_{\bm{\Psi}}$}\\
\midrule
\multirow{4}{*}{0.1}
  & C = 500 & 0.325(0.014) & 0.510(0.018) & 0.314(0.013) & 0.690(0.007) & 0.553(0.009) & 0.919(0.005)\\
  & C = 1000 & 0.309(0.015) & 0.414(0.016) & 0.300(0.016) & 0.793(0.008) & 0.669(0.010) & 0.974(0.003)\\
  & C = 1500 & 0.307(0.016) & 0.391(0.018) & 0.297(0.016) & 0.827(0.013) & 0.711(0.018) & 0.989(0.002)\\
  & C = 2000 & 0.277(0.012) & 0.360(0.012) & 0.288(0.015) & 0.853(0.004) & 0.748(0.007) & 0.994(0.002)\\
\cmidrule(lr){1-8}
\multirow{4}{*}{0.3}
  & C = 500 & 0.393(0.012) & 0.577(0.017) & 0.435(0.016) & 0.764(0.008) & 0.647(0.010) & 0.933(0.008)\\
  & C = 1000 & 0.388(0.018) & 0.519(0.016) & 0.454(0.015) & 0.852(0.006) & 0.756(0.010) & 0.977(0.003)\\
  & C = 1500 & 0.387(0.013) & 0.507(0.018) & 0.456(0.012) & 0.887(0.008) & 0.804(0.013) & 0.989(0.002)\\
  & C = 2000 & 0.366(0.017) & 0.489(0.022) & 0.452(0.018) & 0.908(0.003) & 0.835(0.004) & 0.994(0.002)\\
\cmidrule(lr){1-8}
\multirow{4}{*}{0.5}
  & C = 500 & 0.458(0.014) & 0.618(0.014) & 0.533(0.017) & 0.810(0.017) & 0.712(0.023) & 0.940(0.006)\\
  & C = 1000 & 0.438(0.018) & 0.555(0.013) & 0.555(0.011) & 0.884(0.034) & 0.809(0.053) & 0.975(0.006)\\
  & C = 1500 & 0.442(0.018) & 0.556(0.026) & 0.560(0.011) & 0.919(0.027) & 0.863(0.046) & 0.986(0.004)\\
  & C = 2000 & 0.421(0.019) & 0.526(0.019) & 0.557(0.011) & 0.935(0.030) & 0.886(0.052) & 0.992(0.004)\\
\bottomrule
\end{tabular}%
}
\caption{\small Diffusion network estimations from the proposed method under different degrees of overlapping between diffusion networks.}
\label{tab:overlap}
\end{table}
Table \ref{tab:overlap} shows the diffusion network recovery performance of the proposed method improves as sample size increases at each overlapping degree. With fixed sample size, the MAE of parameter estimations increases while topology recovery performance improves as the degree of overlapping increases. This pattern reveals the trade-off between network differentiation and network topology recovery in the mixture model. As the overlapping level increases, more cascade samples can be mutually borrowed for identifying non-zero edges in either $\bm{\Theta}$ or $\bm{\Psi}$. On the other hand, it becomes more difficult and requires more samples to differentiate $\bm{\Theta}$ and $\bm{\Psi}$ on their overlapped supports. 

\vspace{-5mm}
\subsection{Performance under different time window length}

\begin{figure}[htbp]
\centering
\hspace{-5mm}%
\begin{subfigure}{0.36\textwidth}
  \centering
  \includegraphics[width=\textwidth,height=4cm]{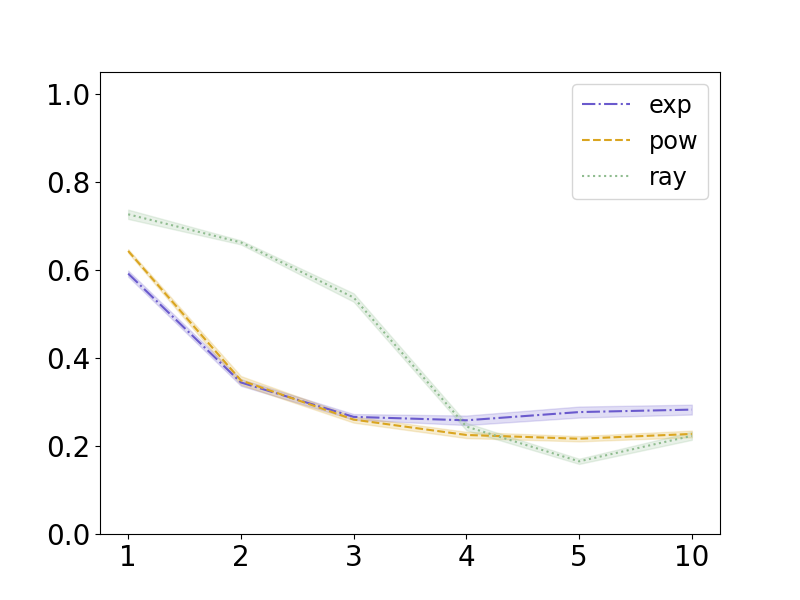}
  \vspace{-5mm}
  \caption{MAE($\bm{\Theta}$)}
\end{subfigure}\hspace{-5mm}
\begin{subfigure}{0.36\textwidth}
  \centering
  \includegraphics[width=\textwidth,height=4cm]{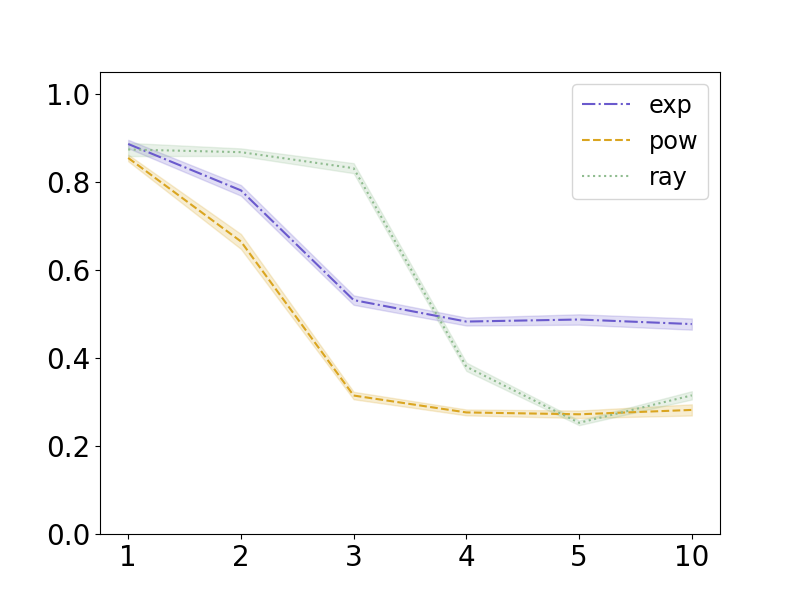}
  \vspace{-5mm}
  \caption{MAE($\bm{\Psi}$)}
\end{subfigure}\hspace{-5mm}
\begin{subfigure}{0.36\textwidth}
  \centering
  \includegraphics[width=\textwidth,height=4cm]{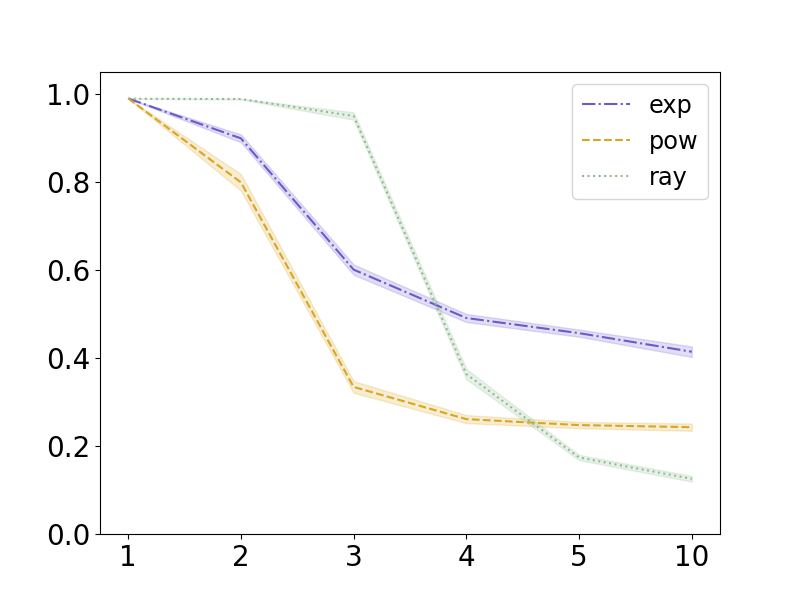}
  \vspace{-5mm}
  \caption{MAE($\bm{\pi}$)}
\end{subfigure}\hspace{-5mm}

\vspace{-5mm}
\caption{\small Parameter estimation under different time window lengths.}
\label{fig:windowmae}
\end{figure}

\begin{figure}[htbp]
\centering
\hspace{-5mm}%
\begin{subfigure}{0.36\textwidth}
  \centering
  \includegraphics[width=\textwidth, height=4cm]{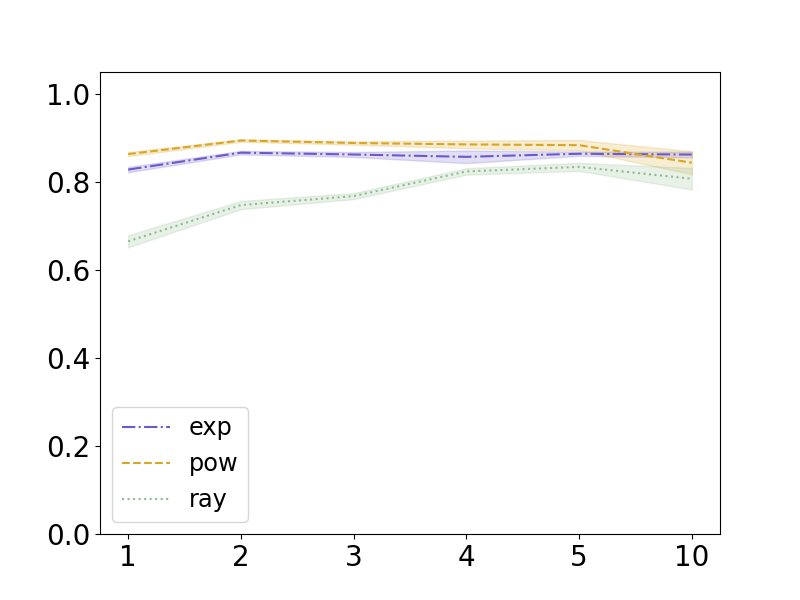}
  \vspace{-5mm}
  \subcaption{Accuracy($\bm{\Psi}$)}
\end{subfigure}\hspace{-5mm}%
\begin{subfigure}{0.36\textwidth}
  \centering
  \includegraphics[width=\textwidth, height=4cm]{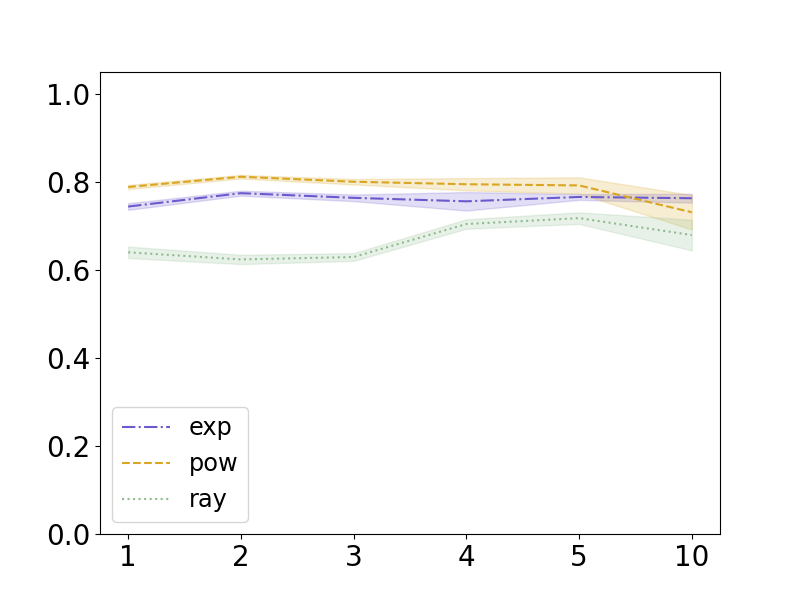}
  \vspace{-5mm}
  \subcaption{Precision($\bm{\Psi}$)}
\end{subfigure}\hspace{-5mm}%
\begin{subfigure}{0.36\textwidth}
  \centering
  \includegraphics[width=\textwidth, height=4cm]{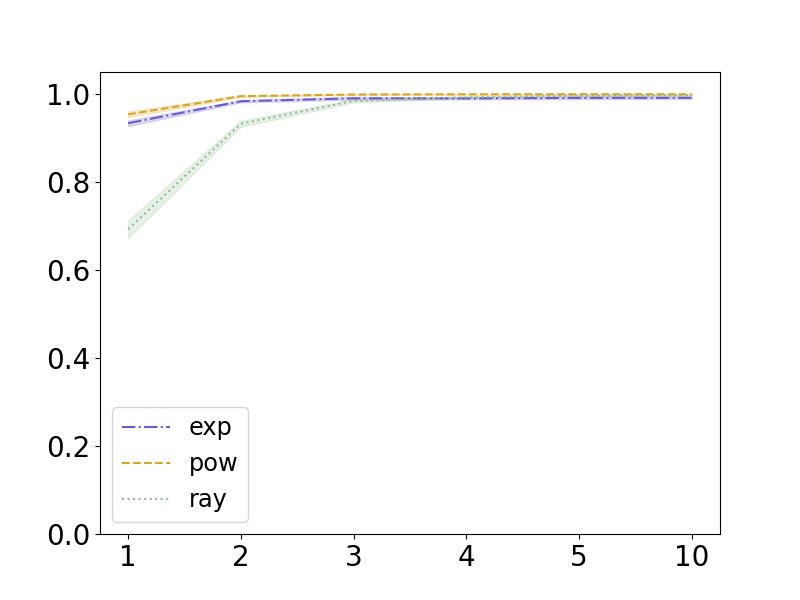}
  \vspace{-5mm}
  \subcaption{Recall($\bm{\Psi}$)}
\end{subfigure}
\hspace{-15mm}%

\vspace{-5mm}
\caption{\small Topology recovery of latent network estimation under different time window lengths. }
\label{fig:windowother}
\end{figure}

In this subsection, we investigate the network estimation performance under different observation window lengths, which ranges in $T \in {\{1, 2, 3, 5, 10\}}$. We generate $\bf{\Theta}$ and $\bf{\Psi}$ following the same procedure in Section 5.3 and fix their overlap level at about $0.30$. We generate $C = 1500$ cascade samples from transmission models Exp, Pow, and Ray, respectively. The parameter estimation accuracies is illustrated in Figure \ref{fig:windowmae} where estimation accuracies improve as the observation window becomes $T$ longer since effective diffusion information increases within each cascade sample. This empirical result is also consistent with our theoretical analysis. Specifically, the proposed method achieves better MAE of $\bm{\Theta}$ and $\bm{\Psi}$ under Ray and Pow transmission models than Exp model.
Figure \ref{fig:windowother} shows that network topology recovery of $\bm{\Psi}$ also improves as $T$ increases and then remains stable. 
In addition to the above experiments, we investigate on the performance of our proposed method when support of diffusion network \bm{$\Theta$} is unobserved. Due to the space limits, we defer the simulation results and discussion in Appendix. 

\vspace{-5mm}

\section{Real cascade data analysis}

In this section, we apply the proposed method to study research topic diffusion networks among universities in the United States. Inferring the knowledge diffusion networks among universities helps us to understand how research ideas spread across institutions and how influential universities shape the research trends. In this application, we focus on the cascading patterns of research topics in sociology. Geographic proximity is known to facilitate idea exchange via collaborations among colleagues within the same institution or nearby locations \citep{balland2017geography}, which can be due to dependence on shared research resources and the need for coordination \citep{morrison2020challenges}. It is also known that the research diffusion can proceed via latent network among scholars, referred as the "invisible colleges" \citep{Crane1972-CRAICD} or "weak ties" \citep{granovetter1973strength}. These concepts highlight the existence and role of informal networks and latent ties, where the connection between idea generators and receivers is often indirect. These latent connections can based on researchers training at similar institutions, attending similar conferences, and following the similar journals, which typically introduce diffusion patterns different than the proximity-based diffusion network. Our goal is to infer both proximity-based diffusion network and latent diffusion network from the research topic cascades. 

To construct research topics, we utilize Elsevier's Scopus API to compile a dataset of 29,725 unique articles from 23 top generalist sociology journals, and extract information from each article including authors with affiliations, article titles, keywords, and abstracts. We then align the article-specific keyphrases to obtain a pool of representative research topics, and assign each article at most five representative research topics. 
To construct the set of universities, we first select universities in both list of sociology graduate programs published by the American Sociological Association in 1965 and list of 2022 U.S. News Best Sociology Programs in United States. We then match the selected universities against the unique affiliation names previously extracted from articles, and finalize $N = 104$ universities. We create the geographical network among these $104$ universities as $\bm{A}\in \{0,1\}^{N \times N}$ such that $\bm{A}_{ij} = 1$ if university $i$ and $j$ are located in the same state. After the pre-processing, we obtain $3,033$ unique research topic cascade samples. For each research topic $c$, the corresponding cascade sample is encoded as $(t_1^c,t_2^c,\cdots, t_{N}^c)$
where $t_i^c: = \tilde{t}_i^c - \tilde{t}_0^c$ with $\tilde{t}_0^c$ being the publication date of the first article which involves topic $c$, and $\tilde{t}_i^c$ being the publication date of the first article which involves topic $c$ and is published by scholar affiliated with university $i$. If university $i$ never publishes an article on the topic $c$, then we set $t_i^c$ as year 2022. Detailed pre-processing of the research topic cascade data are provided in Appendix. 

\begin{figure}[h]
\centering
  \includegraphics[width=0.85\textwidth]{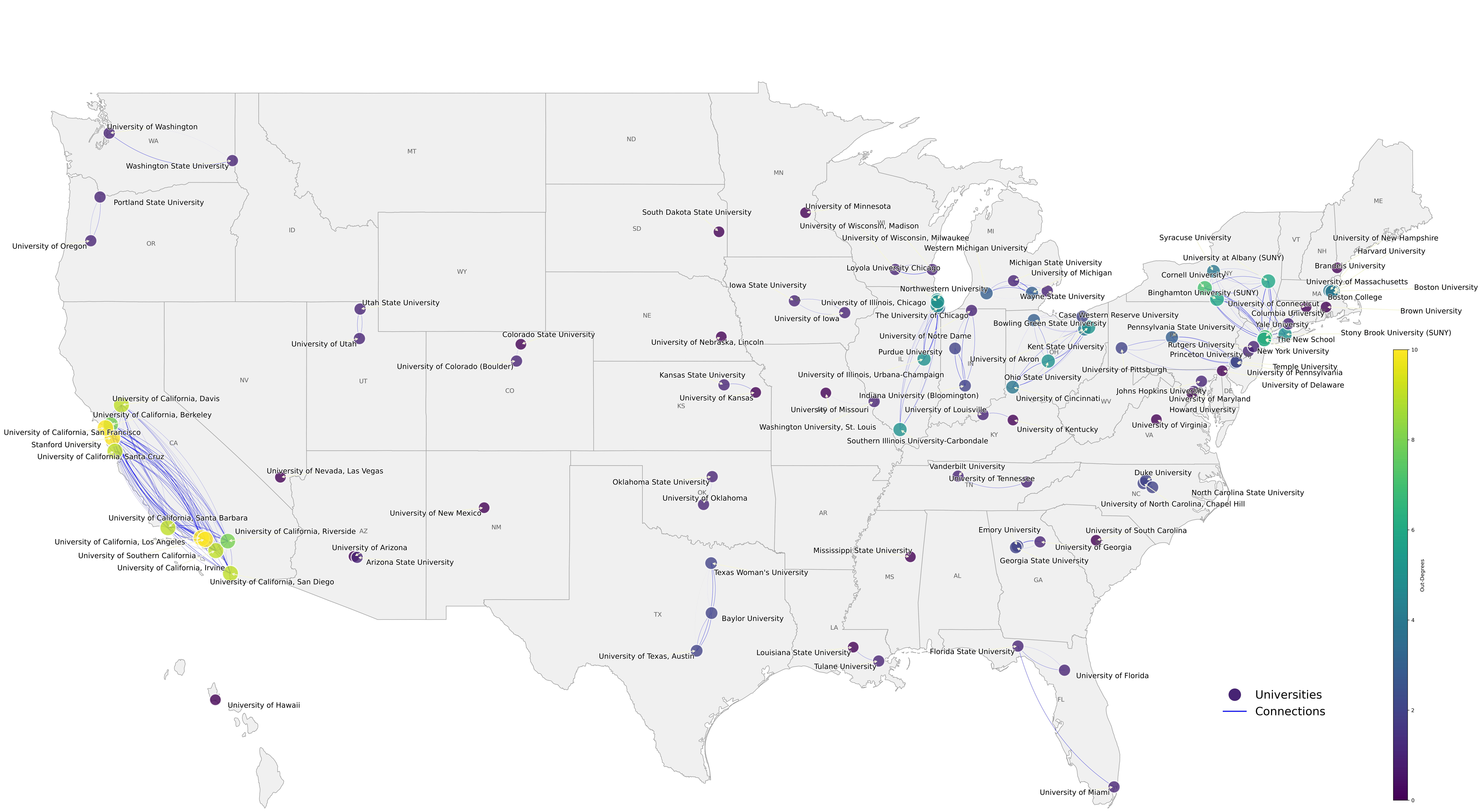}
  \caption{\small Cascade transmission rates over geographical network $\bm{\Theta}$}
\label{fig:geo_network}
\end{figure}

\begin{figure}[h]
\centering
  \includegraphics[width=0.85\textwidth]{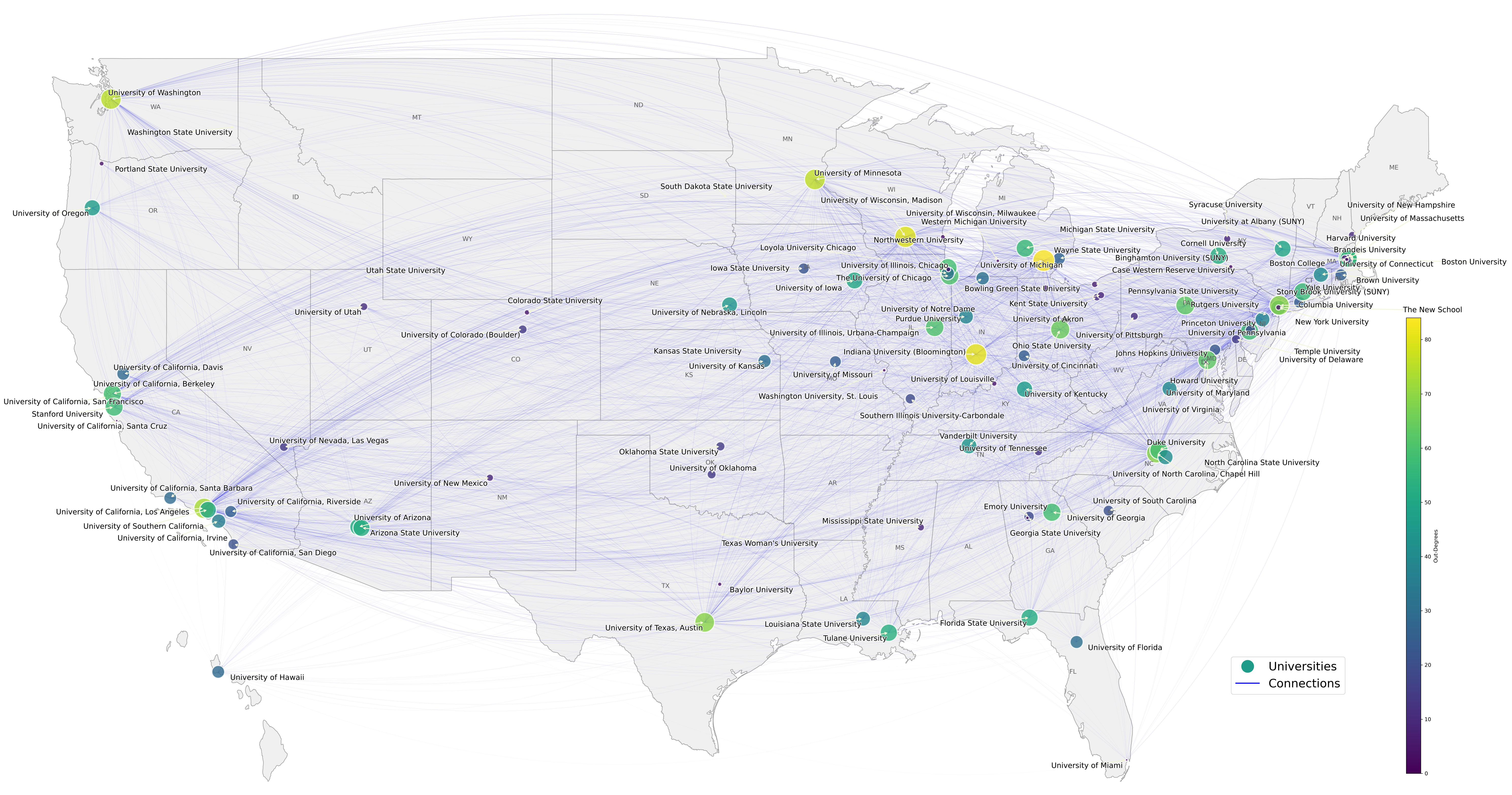}
  \caption{\small Cascade transmission rates over latent diffusion network $\bm{\Psi}$}
\label{fig:latent_network}
\end{figure}

We estimate the geographic diffusion network $\bm{\Theta}$ and latent diffusion network $\bm{\Psi}$ based on proposed double graphic model. Specifically, we randomly select $80\%$ of the total cascade samples as training data, and remaining $20\%$ samples as validation samples for hyperparameter tuning. Figure (\ref{fig:geo_network}) and Figure (\ref{fig:latent_network}) illustrate the inferred diffusion networks $\bm{\hat{\Theta}}$ and $\bm{\hat{\Psi}}$, respectively. The size of nodes represents the out-degree of corresponding university on networks, and the width of line represents the edge-wise transmission rates.  
The geographic diffusion network has a strong local community structure, especially among universities among California, the great lakes, and northeastern coast. On the other hand, the latent diffusion network demonstrates a decentralized structure, and contains many connections between east and west coast, which captures the nation-wide collaboration and academic mobility. 
In addition, we investigate and compare pairwise transmission rates on diffusion networks $\bm{\hat{\Theta}}$ and $\bm{\hat{\Psi}}$ in Figure 9(a). Overall, the latent diffusion network is much more dense than geographic network. Compared with latent diffusion network, the magnitude of transmission rates on geographic network is relatively large and has larger variation.  
\begin{figure}[h]
  \centering
\begin{subfigure}[b]{0.45\textwidth}
    \centering
    \includegraphics[width=\linewidth, height=7cm]{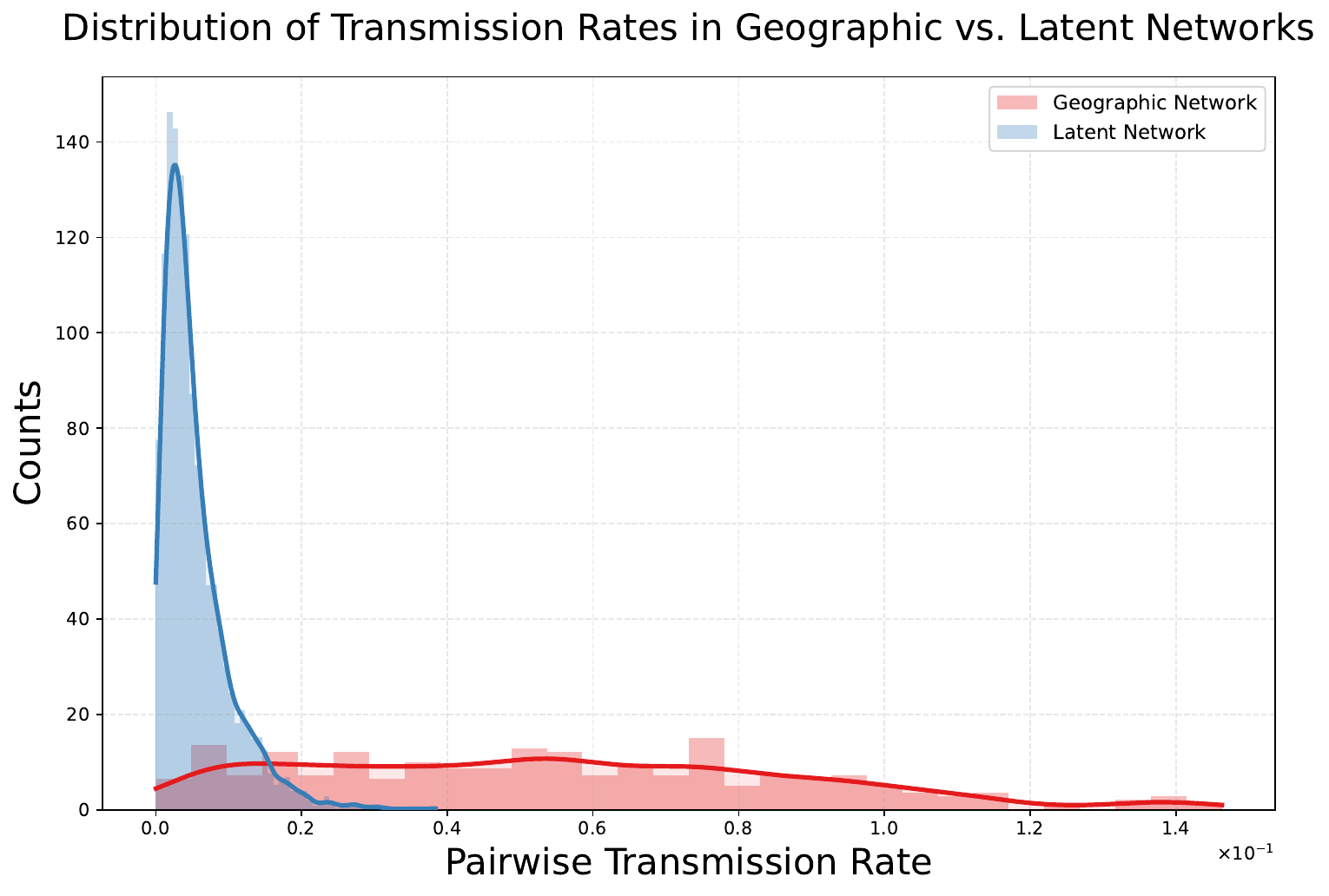}
    \label{fig_prob:grid:b}
  \end{subfigure}
   \begin{subfigure}[b]{0.45\textwidth}
    \centering
    \includegraphics[width=\linewidth, height=5.3cm]{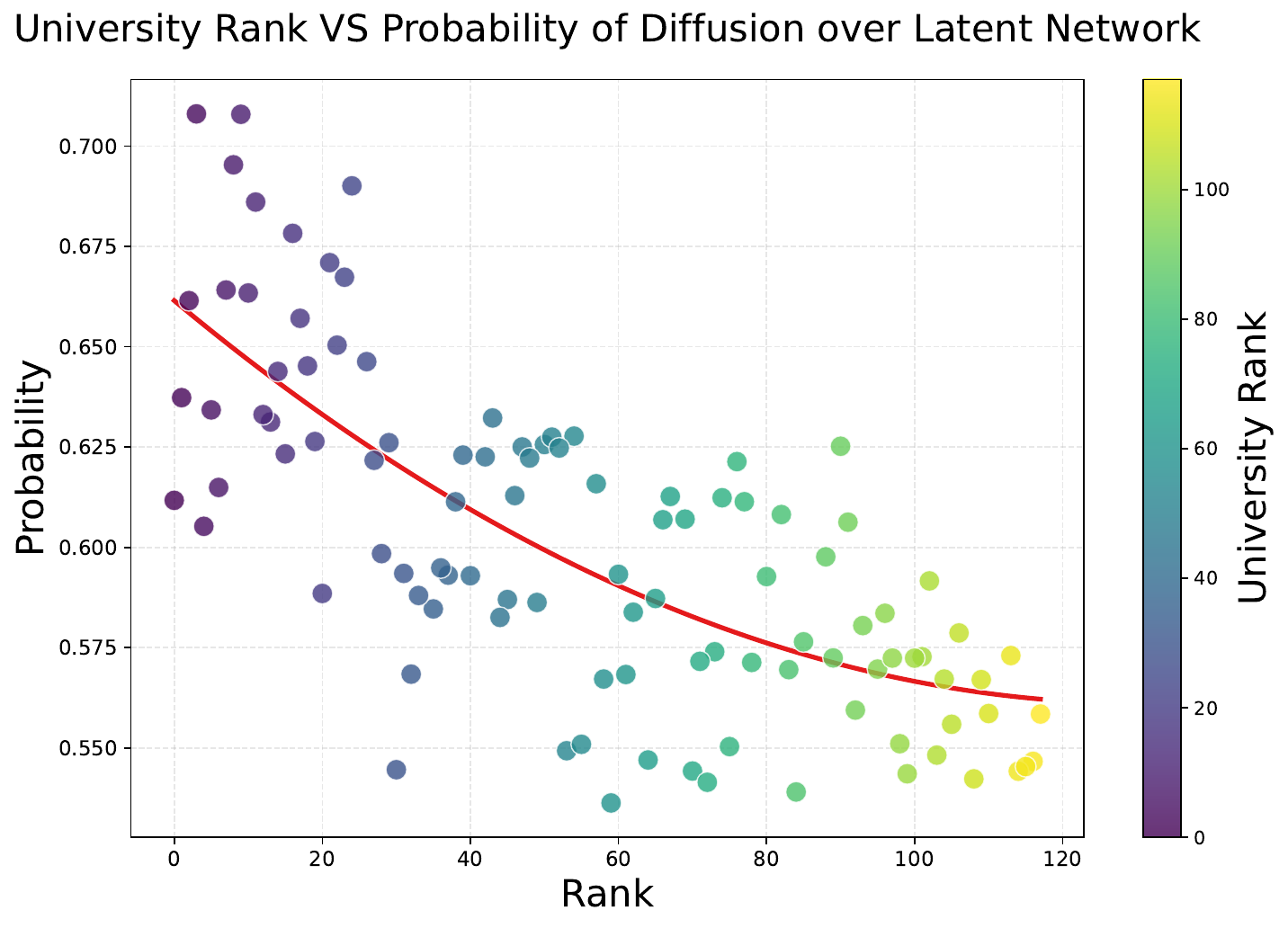}
    \label{fig_prob:grid:a}
  \end{subfigure}
  \vspace{-15mm}
\caption{\small (a) Comparison of distributions of pairwise transmission rates from geographic network $\bm{\Theta}_{ij}$ and latent network $\bm{\Psi}_{ij}$; (b)  Probability of latent network vs Ranking}
  \label{fig:latent_preference}
\end{figure}
To further analyze the inferred geographic diffusion network $\bm{\Theta}$ and latent diffusion network $\bm{\Psi}$, we investigate the relation between universities' positions on the diffusion networks and universities' ranking based on U.S. News University ranking on the sociology program, which is a systematic and popular summary of academic factors. Figure (\ref{fig:degree_vs_rank}) shows the relation between program ranking and out-degree of each university on two diffusion networks. 
\begin{figure}[h]
  \centering
  \begin{subfigure}[b]{0.48\textwidth}
    \centering
    \includegraphics[width=\linewidth]{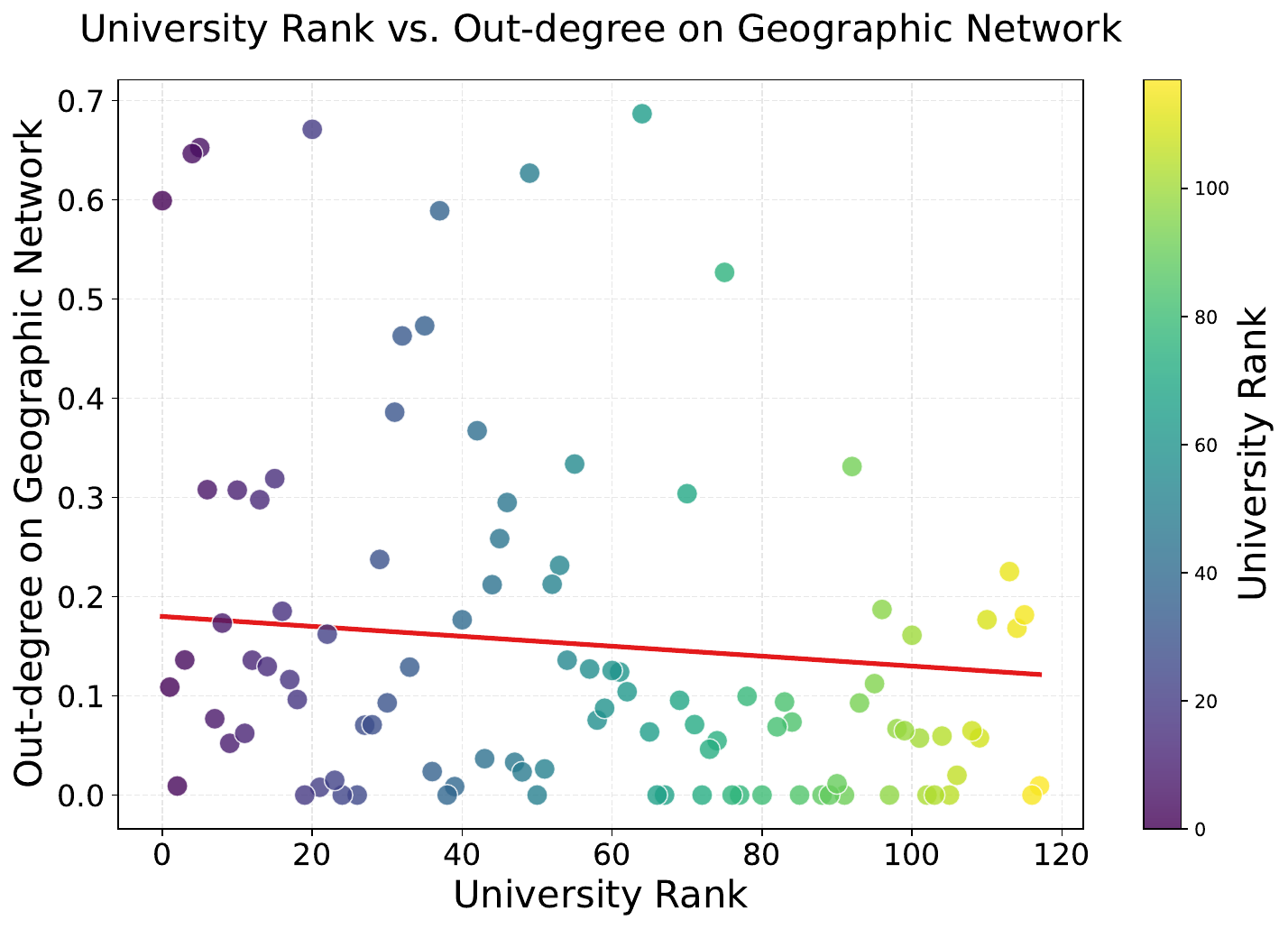}
    \vspace{-12mm}
    \subcaption{\small Outdegree on $\bm{\Theta}$ vs Ranking}
    \label{fig:grid:c}
  \end{subfigure}
  \begin{subfigure}[b]{0.48\textwidth}
    \centering
    \includegraphics[width=\linewidth]{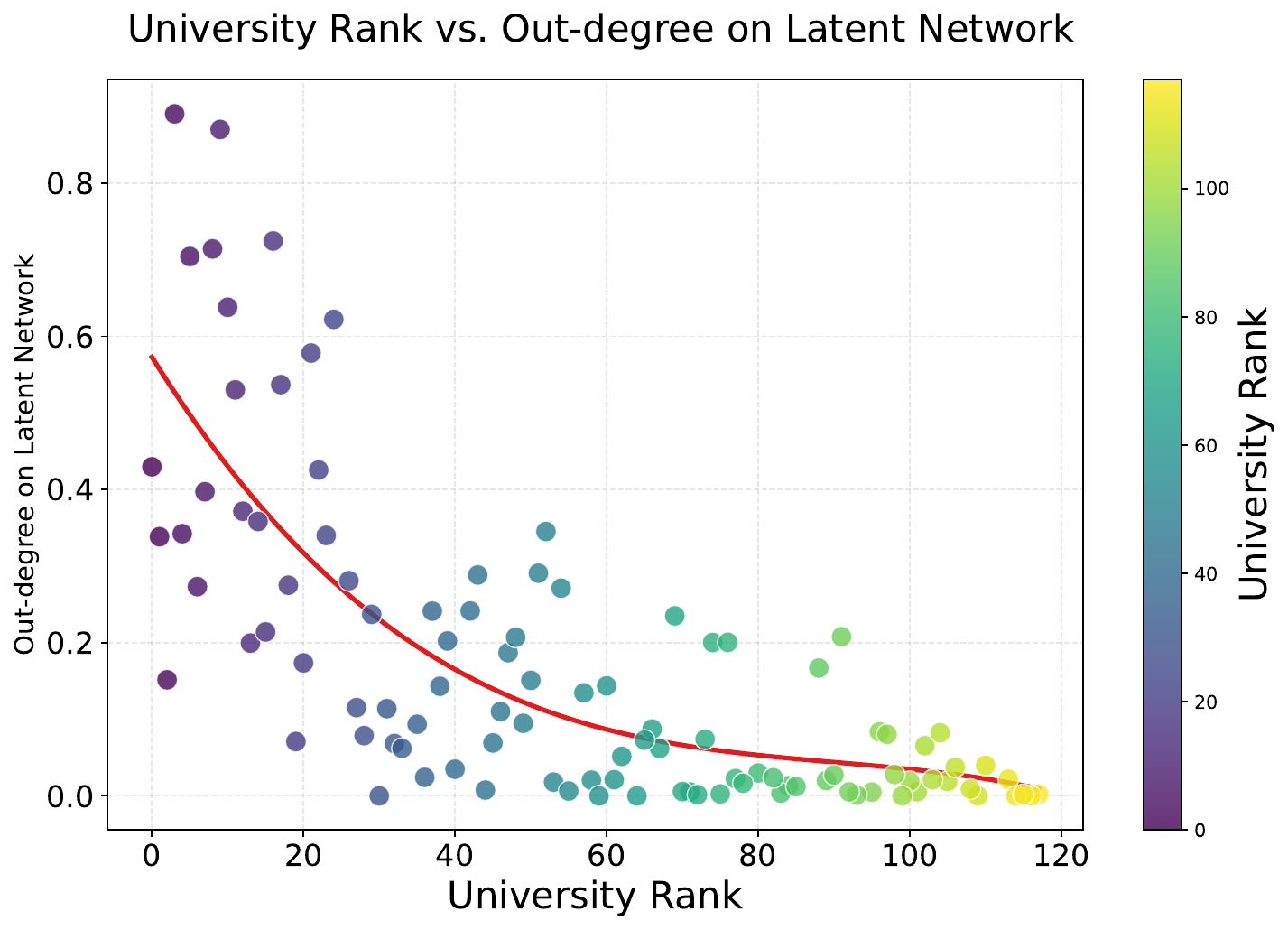}
    \vspace{-12mm}
    \subcaption{\small Outdegree on $\bm{\Psi}$ vs Ranking}
    \label{fig:grid:d}
  \end{subfigure}
  \vspace{-5mm}
  \caption{\small Compare the node-wise outdegree on diffusion networks $\bm{\Theta}$ and $\bm{\Psi}$ with U.S. news sociology program ranking}
  \label{fig:degree_vs_rank}
\end{figure}
The results show that higher ranked universities have higher out-degrees on both diffusion networks in overall. This observation implies that high-ranked universities are active in research productivity, i.e., they published more new ideas (out-degree) 
On the other hand, as universities' ranking becoming higher, node degree increases faster on latent diffusion network than on geographic diffusion network, indicating that the difference on intensity of idea exchange between low-ranked and high-ranked universities is larger on latent network. The diffusion powered by geographic proximity can be limited while more ideas spread by high-ranked universities via latent network. Furthermore, node degree on latent network is more consistent with the ranking compared with geographic network.

\begin{figure}[h]
  \centering
  \begin{subfigure}[b]{0.48\textwidth}
    \centering
    \includegraphics[width=\linewidth]{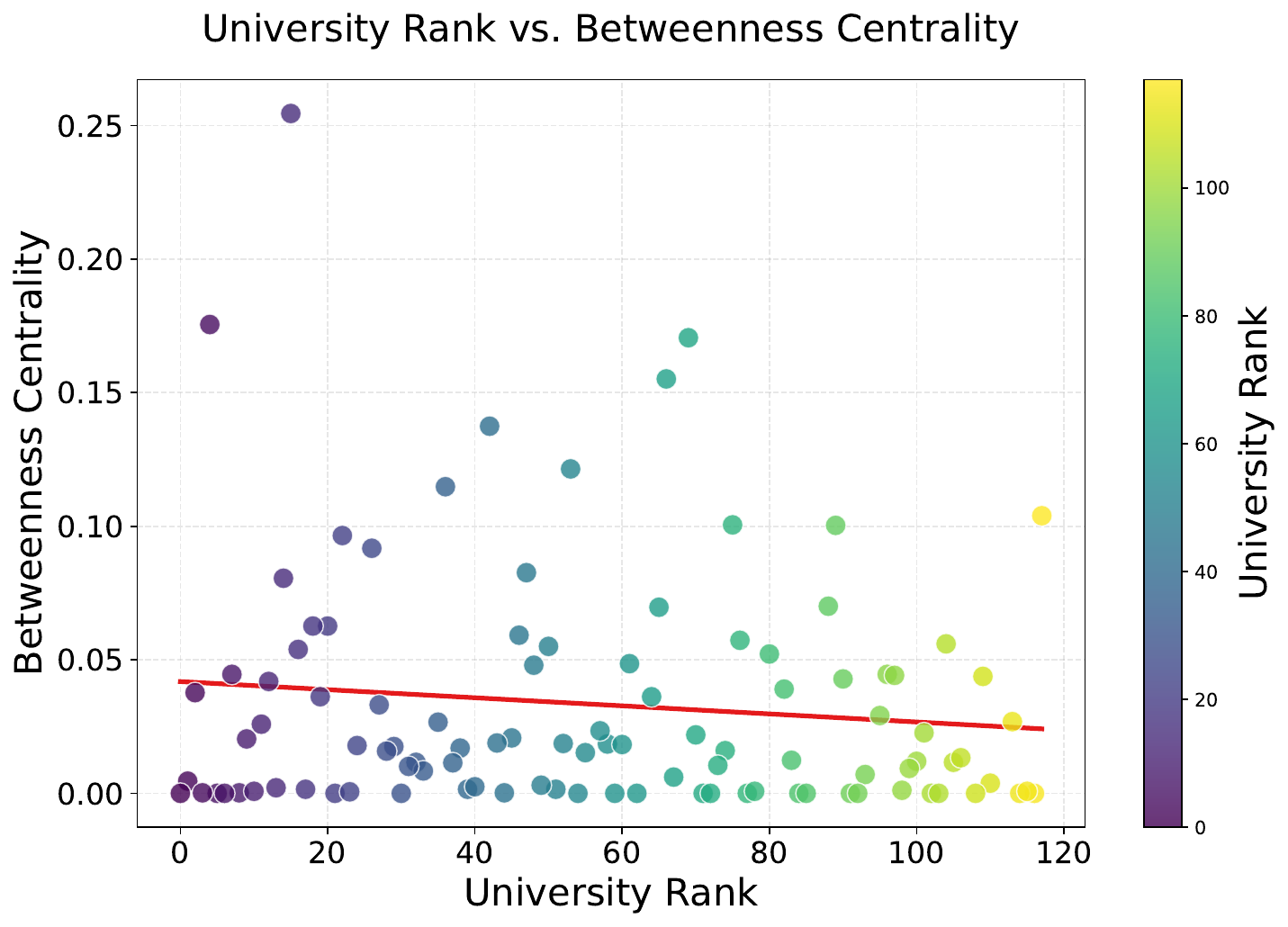}
    \vspace{-12mm}
    \subcaption{Node-wise betweenness centrality on $\bm{\Psi}$}
    \label{fig:grid:a}
  \end{subfigure}
  \begin{subfigure}[b]{0.48\textwidth}
    \centering
    \includegraphics[width=\linewidth]{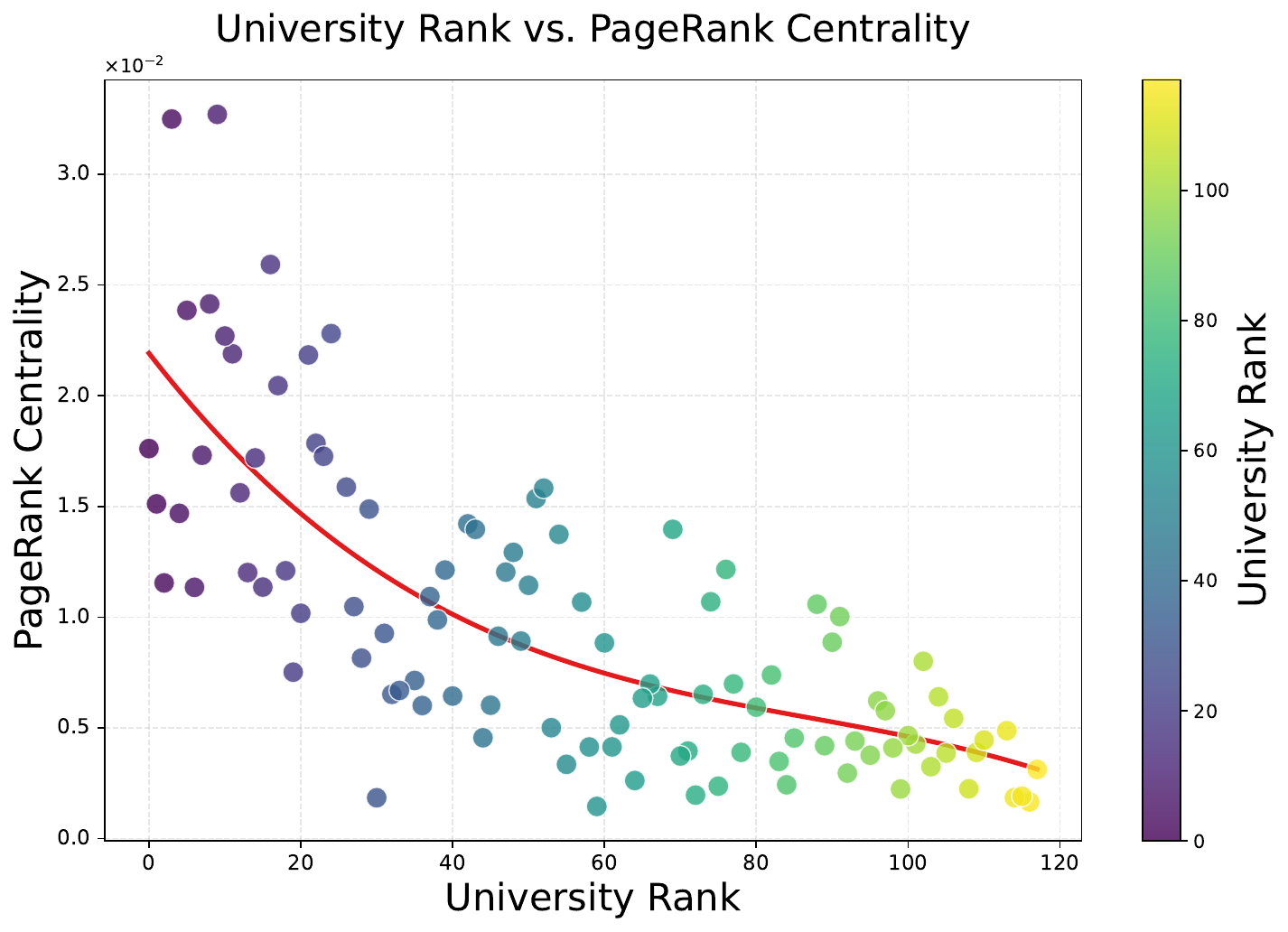}
     \vspace{-12mm}
    \subcaption{Node-wise pagerank centrality on $\bm{\Psi}$}
    \label{fig:grid:b}
  \end{subfigure}
    \vspace{-5mm}          
\caption{\small Node-wise centrality on latent network vs Ranking}
  \label{fig:centrality_vs_rank}
\end{figure}

Figure (\ref{fig:grid:a}) investigates the association of program ranking with node-wise betweenness centrality on latent diffusion network. Betweenness centrality of node $i$ is defined as $\sum_{j \neq i \neq k} \frac{\sigma_{j k}(i)}{\sigma_{j k}}$ where $\sigma_{j k}$ is the total number of shortest paths from node $j$ to $k$, and $\sigma_{j k}(i)$ is the number of those paths passing through $i$. 
The universities' betweenness centralities are relatively uniform and does not significantly decrease as ranking get higher. This suggests there does not exist strong community or centralized topology in latent network. When group universities according to ranking (top 20, middle 60, low 20 etc.), each group has different high-betweenness universities, serving as bridges in idea diffusion. 
We also investigate the relation between rankings and Pagerank centrality. Pagerank centrality measures the influence of a node on the network that depends on how many influential nodes it connected with. Figure (\ref{fig:grid:b}) shows a decay trend of universities' Pagerank centrality in terms of rankings, which suggests that the idea exchange among high-ranked universities is more frequent and faster than other universities. This pattern can be due to that high-ranked universities have a very high level of research activity. Also, institutional prestige might be considered as an important proxy for the quality of ideas, and hence increases the likelihood of ideas from higher ranked universities to diffuse. Furthermore, we estimate the probability of each university engaging topic diffusion via latent network $\{1-\hat{\pi}_i\}_{i=1}^{N}$.. The average of $\{1 - \hat{\pi}_i\}$ over $104$ US universities is $0.601$, indicating that research topics diffuse via latent network more frequently than via geographic network. Specifically, we compare probabilities of diffusing over latent network with rankings in Figure 9(b),
which suggests that high-ranked universities have higher tendency of exchanging idea via latent network.

\vspace{-5mm}
\section{Conclusion}\label{sec-conc}

In this paper, we propose a novel double mixture directed graph model for heterogeneous cascade data. Our method can model the cascading process over multi-layer networks and identify the latent diffusion networks capturing complementary information on diffusion patterns. One main advantage of our model is the convexity in terms of parameters, which enable the proposed optimization method enjoys both statistical and computational guarantees.  
Due to the nature of EM-type algorithm, we acknowledge scalability of the proposed diffusion network estimation can be a limitation, which leave as an important future direction. Another interesting future work is to generalize the mixture graph model to a system with more than two-layer networks and derive the model identification conditions.










\vspace{-5mm}
{ \bibliography{mix_net_ref.bib}}

\end{document}